\documentclass{aamm} 
\usepackage{newpxtext, newpxmath}
\renewcommand\thisnumber{x}
\renewcommand\thisyear {2025}
\renewcommand\thismonth{10}
\renewcommand\thisvolume{xx}

\usepackage{xeCJK}

\usepackage{booktabs}
\usepackage{siunitx} 
\usepackage{subfig}
\usepackage{xcolor}
\usepackage{listings}
\usepackage{multirow}
\usepackage{algorithm}
\usepackage{algpseudocode}

\begin{document}

%%%%% title and author(s):
% \markboth{Author(s)}{Short Title}
% \title{Title}

\markboth{P. Wang et al.}{A High-Performance Solver in FEALPy}
\title{A High-Performance, Cross-Platform Open-Source Solver for the Incompressible Navier-Stokes Equations in FEALPy}

\author[Pengxiang Wang ,Xianbo Huang,Li Peng and Huayi Wei]{Pengxiang Wang\affil{1}, Xianbo Huang\affil{3}, Li Peng\affil{3} and Huayi Wei\affil{2}\comma\corrauth}
\address{
	\affilnum{1}\ 
	School of Mathematics and Computational Science, Xiangtan University, Xiangtan 411105, China\\
	\affilnum{2}\ 
	School of Mathematics and Computational Science, Xiangtan University,Xiangtan, China; 
	National Center of  Applied Mathematics in Hunan, Xiangtan, China; 
	Hunan Key Laboratory for Computation and Simulation in Science and Engineering, Xiangtan, China\\
	\affilnum{3}\ 
	 National-certiﬁed Enterprise Technology Center, Kingfa Science and Technology Co., Ltd., Guangzhou, 510663, China;
}

\emails{{\tt 202231510120@smail.xtu.edu.cn} (P.Wang), {\tt weihuayi@xtu.edu.cn}(H.Wei)
, {\tt pengli1@kingfa.com.cn} (Li Peng), {\tt huangxianbo@kingfa.com.cn} (Xianbo Huang)}
%

%%%%% Begin Abstract %%%%%%%%%%%
\begin{abstract}
To address the dual challenges of performance portability across heterogeneous hardware and the high usability barriers of conventional computational fluid dynamics (CFD) software, this paper introduces FEALPy.CFD, a high performance, open-source solver for the incompressible Navier-Stokes equations developed within the FEALPy framework. The solver's core innovation is its backend-agnostic design, which supports multiple computational backends like NumPy, PyTorch, and JAX to enable seamless switching between CPU and GPU computations with minimal code modification, thereby maximizing hardware utilization and code portability. Its highly modular architecture provides a library of composable components for various spatial discretization schemes, greatly simplifying the development and integration of new algorithms.Validation on benchmark cases confirms that the implemented numerical schemes achieve their theoretical orders of convergence. Furthermore, the capability to select a suitable backend architecture for different computational tasks fully leverages the hardware's potential, delivering substantial efficiency gains. By lowering the technical barrier to high-performance, cross-platform fluid dynamics simulation, FEALPy.CFD offers a powerful and accessible tool for academic research, engineering applications, and reproducible computational science.
\end{abstract}
%%%%% end %%%%%%%%%%%

%%%%% Keywords %%%%%%%%%%%

%%%% AMS subject classifications %%%%

%%%% maketitle %%%%%
\maketitle
\thispagestyle{empty}

%%%% Start %%%%%%
\section{Introduction}
As a core branch of fluid mechanics, CFD has evolved from a fundamental research tool into an indispensable instrument for modern engineering design and scientific discovery\cite{FERZIGER}. It provides profound insights into complex flow phenomena—often inaccessible through traditional experimental methods—by numerically solving the governing equations of fluid motion, the Navier-Stokes (NS) equations . Its applications are widespread, spanning fields from aerospace and automotive design to biomedical engineering.

Open-source solvers for the NS equations, such as OpenFOAM, FEniCS, and deal.II, have achieved widespread adoption in both academia and industry. These platforms offer powerful numerical capabilities that have driven the advancement of CFD technology. However, they also present several challenges.The common challenges of these open-source NS software packages can be summarized as follows.
First, there is a lack of algorithmic completeness and flexibility. The numerical core of existing tools is often entrenched in a single discretization method for instance, OpenFOAM is primarily based on the Finite Volume Method (FVM), whereas FEniCS and deal.II focus on the FEM. This design restricts the ability of researchers to select the optimal numerical method based on the specific characteristics of the physical problem.
Second, they are often characterized by a steep learning curve. OpenFOAM and deal.II require deep expertise in programming and numerical analysis, while the variational form definition in FEniCS poses a significant hurdle for non-specialists， making them difficult for interdisciplinary researchers and beginners to adopt quickly.
Furthermore, their cross-platform capabilities are limited. Most existing tools are not fully optimized for modern hardware architectures (e.g., GPUs, TPUs) or distributed computing environments, hindering their seamless operation across different operating systems and computational platforms ,thereby limiting their potential in high-performance computing . With the rapid development of open-source software, many new scientific computing libraries like NumPy, PyTorch, and JAX have emerged with distinct features, yet existing CFD software often targets specific platforms and lacks the versatility to fully support a diverse range of algorithms and libraries, exacerbating compatibility issues.

To address these challenges, this paper introduces FEALPy.CFD, a high-performance, open-source, and cross-platform solver for the incompressible NS equations built upon the FEALPy platform. FEALPy is an open-source intelligent CAX simulation engine designed to provide an efficient, flexible, and extensible software platform for the field of numerical simulation \cite{FEALPy}. Its primary design goals include maintainability, extensibility, simplicity, usability, efficiency, and intelligence. The architecture ensures high cohesion and low coupling, facilitating continuous refactoring and rapid iteration. Its implementation adheres to principles such as object-oriented programming, array programming, leveled architecture, modularity, and standardized interfaces, with a clear separation between the computational front-end and back-end. Together, these principles ensure that FEALPy is not only powerful and flexible but also easy to use and maintain.

FEALPy.CFD is designed to overcome the limitations of existing open-source CFD tools in terms of algorithmic comprehensiveness, user accessibility, and cross-platform adaptability. Its core innovations include: support for multiple spatial discretization methods such as FEM, FVM, and Smoothed Particle Hydrodynamics (SPH) to handle a variety of flow problems; the ability to seamlessly switch between arbitrary mesh topologies (e.g., triangles, quadrilaterals), 2D/3D problems, and arbitrary-order finite elements with single-line commands; and support for multiple computational backends, including NumPy, PyTorch, and JAX, thereby enabling efficient computation on heterogeneous hardware such as CPUs and GPUs to suit different use cases. Its highly modular architecture not only facilitates integration with existing modules for industrial applications but also significantly reduces the implementation burden on developers, allowing them to focus on algorithmic innovation and complex physical modeling. These features give FEALPy.CFD a distinct advantage in enhancing computational efficiency, flexibility, and scalability, opening up new possibilities for CFD education, research, and engineering applications.

The remainder of this paper is structured as follows. 
Section 2 outlines the mathematical models, discretization methods, and solution algorithms used in FEALPy.CFD. 
Section 3 will elaborate on the software architecture of FEALPy.CFD, detailing the functionality and the design rationale behind each of its constituent modules. 
Section 4 provides a demonstration of the installation and application of FEALPy.CFD using a steady-state NS problem as an example. 
Section 5, we conduct a series of numerical experiments to highlight the powerful flexibility of FEALPy.CFD. These examples not only showcase its practical utility in diverse and complex scenarios but also validate the correctness of our schemes by examining their convergence orders. 
Finally, Section 6 summarizes the paper and offers suggestions for future research.

\section{Models and Solution Algorithms}
This section aims to provide a concise overview of the core numerical algorithms implemented in FEALPy.CFD for solving incompressible flow problems. FEALPy.CFD features a comprehensive algorithmic framework, integrating mainstream discretization methods such as the FEM, FVM, and SPH. This framework enables the efficient solution of various governing equations, including the Stokes, steady-state, and unsteady NS equations.

Due to space constraints, this chapter will not delve into exhaustive theoretical derivations. We begin by introducing the governing equations. Subsequently, we will present the numerical algorithms, organized according to their spatial discretization methods implemented in FEALPy.CFD.

\subsection{Governing Equations}
This part introduces the governing equations currently supported by the FEALPy.CFD framework. All models are based on the principles of incompressible fluid dynamics, aiming to solve for the velocity field $\boldsymbol u$ and the pressure field $p$ within a given computational domain$\Omega$.

The first set of equations considered is the Stokes equations, which describe "creeping flow"—a regime where viscous forces dominate inertial forces. This corresponds to very low Reynolds numbers $(Re \ll 1)$. The equations are formulated as follows:

\begin{equation}
	\begin{cases} \label{Stokes_equation}
		-\nu\Delta\boldsymbol{u}+\nabla p=\boldsymbol{f} & \mathrm{in}\Omega, \\
		\nabla\cdot\boldsymbol{u}=0 & \mathrm{in}\Omega.  
	\end{cases}
\end{equation}

In \eqref{Stokes_equation} $\nu$ represents the kinematic viscosity of the fluid, and $\boldsymbol f$ is the body force term. The most significant features of this equation system are its linearity and the strong coupling between the velocity and pressure fields. Consequently, the primary numerical challenge lies in efficiently and stably solving the resulting saddle-point system, while simultaneously avoiding spurious pressure solutions.

For flows at higher Reynolds numbers, the inertial effects become non-negligible and must be considered. In this regime, the flow is described by the steady-state NS equations:

\begin{equation} \label{steady_ns_equation}
	\begin{cases}
		-\nu\Delta\boldsymbol{u}+(\boldsymbol{u}\cdot\nabla)\boldsymbol{u}+\nabla p=\boldsymbol{f} & \mathrm{in}\Omega, \\
		\nabla\cdot\boldsymbol{u}=0 & \mathrm{in}\Omega.  
	\end{cases}
\end{equation}

Compared to the \eqref{Stokes_equation}, the primary distinction of the \eqref{steady_ns_equation} is the inclusion of the nonlinear convective term $(\boldsymbol{u}\cdot\nabla)\boldsymbol{u}$ . The presence of this term renders the entire system of equations nonlinear, precluding a direct solution. Consequently, efficient iterative methods are required to handle this nonlinearity.

For time-dependent flow phenomena, the system must be described by the  NS equations:

\begin{equation}\label{ns_equation}
	\begin{cases}
		\frac{\partial\boldsymbol{u}}{\partial t}-\nu\Delta\boldsymbol{u}+(\boldsymbol{u}\cdot\nabla)\boldsymbol{u}+\nabla p=\boldsymbol{f} & \mathrm{in}\Omega, \\
		\nabla\cdot\boldsymbol{u}=0 & \mathrm{in}\Omega.  
	\end{cases}
\end{equation}

This system extends the steady-state formulation by including the transient term $\frac{\partial\boldsymbol{u}}{\partial t}$. Its inclusion introduces a new numerical challenge: the time dimension must now be discretized, and the solution advanced forward in time at each step while preserving accuracy. Consequently, two aspects become critically important: first, the use of efficient and stable time-integration schemes; and second, the availability of fast solvers for the coupled velocity-pressure system that must be resolved within each time step.

However, in numerous applications, such as polymer processing, fluids exhibit complex non-Newtonian behaviors. For these materials, the viscosity is not constant but is instead a function of the local shear rate and temperature. To accommodate such materials, FEALPy.CFD provides a flexible interface for various constitutive equations under the Generalized Newtonian Fluid model framework. The current implementation includes two classic models: the Power-Law model \cite{OSTWALD}, which is widely used to describe shear-thinning and shear-thickening characteristics, and the Cross-WLF model \cite{CROSS}. The latter is crucial for applications like injection molding, as it precisely captures the complex viscosity behavior of polymer melts by considering the coupled effects of shear rate and temperature.

\subsection{Numerical Methods}
This part provides a concise overview of the core numerical methods implemented in the FEALPy.CFD framework, which collectively form the computational foundation for tackling a diverse range of fluid dynamics problems. 

\subsubsection{The Finite Element Method}
The core idea of FEM is rooted in variational principles, where the governing differential equation is transformed into its equivalent weak form. When applied to the NS equations, both the FEM and FEM must confront two primary numerical challenges: the nonlinearity arising from the convective term, and the intrinsic coupling of the velocity and pressure fields. To address these, two main classes of strategies are employed: iterative methods are used to handle the nonlinearity, while projection methods are a popular choice for decoupling the velocity and pressure. In the following subsections, building upon the arbitrary-order finite element spaces available in FEALPy.CFD, we will briefly outline the general iterative schemes for linearization  and the fundamental steps of projection methods.

Let $\Omega \subset \mathbb{R}^d $(with d=2 or 3) be a bounded computational domain, which is partitioned by a conforming mesh $\mathcal{T}_h$，$h$denote the mesh size, representing the maximum element diameter in $\mathcal{T}_h$.

The discrete space for the velocity $\mathbf{V}_h$ is defined as:

\begin{equation}
	\mathbf{V}_h=\{\mathbf{v}_h\in[H_0^1(\Omega)]^d\mid\mathbf{v}_h|_K\in[P_k(K)]^d,\forall K\in\mathcal{T}_h\}
\end{equation}

The corresponding discrete space for the pressure $Q_h$ is defined as:

\begin{equation}
	Q_h=\{q_h\in L^2(\Omega)\mid q_h|_K\in P_{k-1}(K),\forall K\in\mathcal{T}_h\}
\end{equation}

where $P_k(K)$denotes the space of continuous Lagrange polynomials of degree up to$k$
on an element $K$.

The idea of linearizing the NS equations by treating the convective velocity as a known quantity was first proposed by Oseen \cite{oseen} . This approach, which leads to the Oseen equations, became one of the early methods for handling the nonlinear convective term

\begin{equation}
	\begin{aligned}
		& \left(\frac{\mathbf{u}_h^{n+1,k+1}-\mathbf{u}_h^n}{\Delta t},\mathbf{v}_h\right)+\left((\mathbf{u}_h^{n+1,k}\cdot\nabla)\mathbf{u}_h^{n+1,k},\mathbf{v}_h\right) \\
		& \qquad +\nu(\nabla\mathbf{u}_h^{n+1,k+1},\nabla\mathbf{v}_h) -\left(p_h^{n+1,k+1},\nabla\cdot\mathbf{v}_h\right)=(\mathbf{f}^{n+1},\mathbf{v}_h),\\
		& \left(\nabla\cdot\mathbf{u}_h^{n+1,k+1},q_h\right)=0,
	\end{aligned}
\end{equation}

A more advanced approach to linearizing the nonlinear system is to use Newton's method. The iterative procedure is as follows:

\begin{equation}
	\begin{aligned}
	\left(\frac{{\bf u}_{h}^{n+1,k+1}-{\bf u}_{h}^{n}}{\Delta t},{\bf v}_{h}\right)
	&+ \left(({\bf u}_{h}^{n+1,k}\cdot\nabla){\bf u}_{h}^{n+1,k+1},{\bf v}_{h}\right) \\
	&+ \left(({\bf u}_{h}^{n+1,k+1}\cdot\nabla){\bf u}_{h}^{n+1,k},{\bf v}_{h}\right)
	- \left(({\bf u}_{h}^{n+1,k}\cdot\nabla){\bf u}_{h}^{n+1,k},{\bf v}_{h}\right) \\
	&+ \nu\left(\nabla{\bf u}_{h}^{n+1,k+1},\nabla{\bf v}_{h}\right)
	- \left(p_{h}^{n+1,k+1},\nabla\cdot{\bf v}_{h}\right)
	= \left({\bf f}^{n+1},{\bf v}_{h}\right), \\
	&\left(\nabla\cdot{\bf u}_{h}^{n+1,k+1},q_{h}\right)
	= 0.
	\end{aligned}
\end{equation}

Another primary strategy for solving the NS equations is the use of projection methods, which significantly reduce computational cost by decoupling the velocity and pressure fields. 

The classic projection method was pioneered independently by Chorin \cite{CHORIN} and Temam \cite{TEMAM}. The core of this method is a fractional-step scheme that first computes an intermediate velocity field $\mathbf{u}^*$ by solving the momentum equation without the pressure gradient. As this intermediate field is not divergence-free, a second step enforces the incompressibility constraint by solving a Poisson equation for the pressure. The resulting pressure is then used to correct the velocity field. The steps are outlined as follows:
\begin{equation}
	\begin{aligned}
	\left(\frac{\mathbf{u}_h^*-\mathbf{u}_h^n}{\Delta t},\mathbf{v}_h\right)+((\mathbf{u}_h^n\cdot\nabla)\mathbf{u}_h^n,\mathbf{v}_h)+\nu(\nabla\mathbf{u}_h^*,\nabla\mathbf{v}_h)&=(\mathbf{f}^{n+1},\mathbf{v}_h),\quad\forall\mathbf{v}_h\in\mathbf{V}_h,\\
	\left(\nabla p_h^{n+1},\nabla q_h\right)&=\frac{\rho}{\Delta t}\left(\nabla\cdot\mathbf{u}_h^*,q_h\right),\quad\forall q_h\in Q_h,\\
	\left(\frac{\mathbf{u}_h^{n+1}-\mathbf{u}_h^*}{\Delta t},	\mathbf{v}_h\right)&=-\frac{1}{\rho}\left(\nabla p_h^{n+1},\mathbf{v}_h\right),\quad\forall\mathbf{v}_h\in\mathbf{V}_h .
	\end{aligned}
\end{equation}

The Incremental Pressure Correction Scheme (IPCS) \cite{GODA} is a refinement of Chorin's method that achieves higher numerical accuracy by introducing an updated treatment of the pressure term. The algorithmic procedure is outlined as follows:

\begin{equation}
	\begin{aligned}
\left(\frac{{\mathbf{u}}_{h}^{*}-{\mathbf{u}}_{h}^{n}}{\Delta t},\mathbf{v}_{h}\right)+(({\mathbf{u}}_{h}^{n}\cdot\nabla){\mathbf{u}}_{h}^{n},\mathbf{v}_{h})+\nu(\nabla{\mathbf{u}}_{h}^{*},\nabla\mathbf{v}_{h})&-(p_{h}^{n},\nabla\cdot\mathbf{v}_{h})=(\mathbf{f}^{n+1},\mathbf{v}_{h}),\quad\forall\mathbf{v}_{h}\in\mathbf{V}_{h},\\
\left(\nabla(p_{h}^{n+1}-p_{h}^{n}),\nabla q_{h}\right)&=\frac{\rho}{\Delta t}(\nabla\cdot{\mathbf{u}}_{h}^{*},q_{h}),\quad\forall q_{h}\in Q_{h},\\
\left(\frac{{\mathbf{u}}_{h}^{n+1}-{\mathbf{u}}_{h}^{*}}{\Delta t},\mathbf{v}_{h}\right)&=-\frac{1}{\rho}\left(\nabla(p_{h}^{n+1}-p_{h}^{n}),\mathbf{v}_{h}\right),\quad\forall\mathbf{v}_{h}\in\mathbf{V}_{h}.
	\end{aligned}
\end{equation}

\subsubsection{The Finite Volume Method}
The FVM is founded upon the integral form of the governing conservation laws. In this method, the computational domain is partitioned into a finite set of non-overlapping control volumes.To prevent the pressure checkerboarding problem that arises on collocated grids, one can either employ a staggered grid arrangement or utilize a special interpolation scheme for the face velocities such as Rhie-Chow interpolation method \cite{CHOW} to correct the velocity at the control volume faces:

\begin{equation} \label{Rhie-Chow}
	\boldsymbol u_f = \overline{\boldsymbol u_f}- D_f\left(\nabla p_f-\overline{\nabla p_f}\right)
\end{equation}

In \eqref{Rhie-Chow} $\overline{\boldsymbol u_f}$ is the face velocity obtained through standard momentum interpolation from the neighboring cell centers.$ D_f$is a coefficient interpolated to the face f. It is derived from the momentum equation coefficients ${D}_P$,of the two cells adjacent to the face f. The coefficient ${D}_P$ is itself related to the diagonal entries of the discretized momentum matrix.$\nabla p_f$ represents the pressure gradient at the face, computed directly by differencing the pressure values at the two adjacent cell centers.$\overline{\nabla p_f}$is an interpolated pressure gradient at the face, obtained by first computing the cell-centered gradients and then averaging them to the face.

FEALPy.CFD provides support for both coupled and decoupled solution strategies. For fully coupled approaches, the aforementioned iterative methods, such as the Newton and Oseen iterations, can be employed to solve the monolithic system.

For decoupled approaches, the framework implements the widely-used SIMPLE and PISO algorithms. The SIMPLE (Semi-Implicit Method for Pressure Linked Equations) algorithm is primarily suited for steady-state problems, while the PISO (Pressure-Implicit with Splitting of Operators) algorithm is designed for transient simulations.

The SIMPLE algorithm, introduced by Patankar and Spalding \cite{PATANKAR}, is an iterative procedure for coupling pressure and velocity. The core idea is a guess-and-correct approach. First, a provisional velocity field $\boldsymbol{u}^*_h$, is calculated by solving the momentum equations using the pressure field from the previous iteration $p^{k}_h$ .Based on this provisional velocity, a pressure-correction equation is then constructed and solved for a pressure correction term $p_h'$ . Subsequently, this pressure correction is used to update the pressure field (and also the velocity field to satisfy continuity). This iterative process is repeated until convergence is reached. The steps of the algorithm are as follows:

\begin{equation}\label{simple_1}
	\begin{gathered}
		\oint_{\partial\Omega_h} \mathbf{u}_h^{*} (\mathbf{u}_h^{k} \cdot \mathbf{n}) dS + \nu \oint_{\partial\Omega_h} \nabla \mathbf{u}_h^{*} \cdot \mathbf{n} dS \mathbf{n} dS =\int_{\Omega_h}\nabla p_h^{k}dV+\int_{\Omega_h} \mathbf{f} dV 
	\end{gathered}
\end{equation}

\begin{equation}\label{simple_2}
	\oint_{\partial\Omega_h} {D}_P \nabla p_h' \cdot \mathbf{n} dS = \int_{\Omega_h} \nabla \cdot \mathbf{u}^* dV
\end{equation}

\begin{equation}\label{simple_3}
	\begin{aligned}\mathbf{u}_h^{n+1} &= \mathbf{u}_h^{*} - {D}_P \nabla p_h'\\
		p_h^{k+1} &= p_h^{k} + \alpha_p p_h'\end{aligned}
\end{equation}

The preceding iterative steps are then repeated until the pressure solution from two consecutive iterations satisfies the condition $\|p_h^{k+1} - p_h^k\| < \epsilon$ , at which point the iteration is terminated.

The PISO  algorithm, developed by Issa \cite{ISSA}, is an extension of the SIMPLE algorithm designed specifically for transient simulations. Its core idea is to achieve a more rigorous pressure-velocity coupling by performing multiple pressure correction steps (typically two) within a single time step. This allows the pressure field to converge more rapidly to the correct solution, thereby avoiding the need for the costly inner iterations that are characteristic of the SIMPLE algorithm in each time step. This approach significantly improves the computational efficiency for unsteady problems. The steps of the algorithm are as follows: 

\begin{equation} \label{piso_1}
	\begin{gathered}
		\int_{\Omega_h}\frac{\mathbf{u}_h^{*}-\mathbf{u}_h^n}{\Delta t}dV+\oint_{\partial\Omega_h} \mathbf{u}_h^{*} (\mathbf{u}_h^{n} \cdot \mathbf{n}) dS + \nu \oint_{\partial\Omega_h} \nabla \mathbf{u}_h^{*} \cdot \mathbf{n} dS =\int_{\Omega_h}\nabla p_h^{n}dV+\int_{\Omega_h} \mathbf{f}^{n+1} dV 
	\end{gathered}
\end{equation}

\begin{equation} \label{piso_2}
	\oint_{\partial\Omega_h} {D}_P \nabla p_h' \cdot \mathbf{n} dS = \int_{\Omega_h} \nabla \cdot \mathbf{u}^* dV
\end{equation}

\begin{equation} \label{piso_3}
	\begin{aligned}
		\mathbf{u}_h^{**} &= \mathbf{u}_h^{*} - {D}_P \nabla p_h'\\
		p_h^{**} &= p_h^{n} + p_h'
	\end{aligned}
\end{equation}

\begin{equation} \label{piso_4}
	\oint_{\partial\Omega_h} {D}_P \nabla p_h'' \cdot \mathbf{n} dS = \int_{\Omega_h} \nabla \cdot \mathbf{u}^{**} dV
\end{equation}

\begin{equation} \label{piso_5}
	\begin{aligned}
		\mathbf{u}_h^{n+1} &= \mathbf{u}_h^{**} - {D}_P \nabla p_h''\\
		p_h^{n+1} &= p_h^{**} + p_h''
	\end{aligned}
\end{equation}

The algorithm then advances to the next time step, and the entire procedure is repeated until the prescribed final simulation time is reached.

\subsubsection{The Smoothed Particle Hydrodynamics Method}

The fundamental principle of SPH is derived from the following integral identity:
\begin{equation}
	f(\boldsymbol{x})=\int_\Omega f(\boldsymbol{y})\delta(|\boldsymbol  r|)\mathrm{d}y,
\end{equation}
In this identity, $f(\boldsymbol x)$ represents an arbitrary field function defined in the domain $\Omega$, the position vectors $\boldsymbol x$ and $\boldsymbol y$ denote two points in $\mathbb{R}^d$, $r = |\boldsymbol x - \boldsymbol y|$ is the distance between them. The core of the SPH method lies in replacing the Dirac delta function$\delta(r)$ with a smoothing kernel function, This kernel approximation yields the following integral representation of the function $f(\boldsymbol{x})$

\begin{equation}
	f(\boldsymbol{x})=\int_\Omega f(\boldsymbol{y})W(|\boldsymbol{r}|,h)\mathrm{d}y,
\end{equation}

Here, the Dirac delta function $\delta(|\boldsymbol{r}|)$ is approximated by a smoothing kernel function $W(|\boldsymbol{r}|,h)$ ,which has a finite support radius defined by the smoothing length $h$. Through this kernel approximation, the SPH algorithm transforms the original continuous integral into a discrete summation over neighboring particles. As a result, any property of a given particle is computed as a weighted average of the same property of its surrounding particles.

\begin{equation}
	f(\boldsymbol{x}_i)=\sum_{j\in\Omega_i}f(\boldsymbol{x}_j)W(|\boldsymbol{x}_i-\boldsymbol{x}_j|,h)V_j.
\end{equation}

where $V_j$ represents the volume associated with particle j. 

The computation of physical quantities and their gradients in the SPH algorithm can be interpreted as a sophisticated interpolation scheme. Through the process of kernel-weighted averaging, information is exchanged among neighboring particles, allowing the discrete particle data to approximate a continuous field. The choice of the kernel function is critical, as it directly influences the accuracy and stability of the numerical simulation. The FEALPy.CFD framework provides a modular library of several widely-used kernel functions. These include standard B-spline kernels such as the Cubic Spline Kernel, Quadratic Kernel, and Quintic Kernel. Additionally, the framework implements advanced kernels from the Wendland family, including the Wendland C2 Kernel and Quintic Wendland Kernel, which are known for their excellent stability properties. This variety enables users to select the optimal kernel to balance computational efficiency, numerical stability, and accuracy for diverse SPH applications.

To address the well-known deficiencies of the classical SPH method—namely its limitations in accuracy, boundary condition enforcement, and numerical stability—the FEALPy.CFD framework incorporates a suite of advanced algorithmic corrections drawn from the literature:
\begin{itemize}
	\item \textbf{Solid Wall Boundary Treatment}\cite{XU_Extension} \cite{REN}: The enforcement of solid wall boundary conditions is handled using the dummy particle method (often synonymous with ghost particles). This technique involves placing several layers of virtual particles on the exterior side of the solid boundary. The physical properties of these dummy particles, such as velocity and pressure, are then determined by extrapolating from the properties of nearby fluid particles in a manner that satisfies the desired boundary condition (e.g., no-slip or free-slip). This approach provides the necessary kernel support near the boundary and effectively prevents fluid particles from unphysically penetrating the solid wall.

	\item \textbf{Kernel Gradient Correction}\cite{XU_Modeling} \cite{BONET} ：To enhance the accuracy of gradient approximations, particularly for particles with irregular or truncated support domains (e.g., those near boundaries), the framework implements a kernel gradient correction method. This technique involves constructing a correction matrix to modify the standard kernel gradient. The application of this correction is crucial as it restores first-order consistency to the gradient approximation, ensuring that linear functions can be reproduced exactly, even in regions where the particle distribution is not uniform.
	
	\item \textbf{Numerical Stability Enhancement} :To suppress the non-physical pressure oscillations that can arise in high-Mach-number flows or scenarios with strong shocks, the framework integrates a classic artificial viscosity model \cite{HE}. Furthermore, for the specific challenges posed by highly viscous fluids (such as polymer melts in injection molding), an improved, low-dissipation Riemann solver is implemented to achieve a more stable and accurate pressure field\cite{REN}.
	
	\item \textbf{Particle Shifting}\cite{REN}\cite{XU_Accuracy} \cite{XU_Technique} :When simulating problems involving large deformations, such as free-surface flows, SPH particles can suffer from unphysical clustering or the formation of voids. To counteract this, the framework employs a particle shifting technique. This algorithm applies a small, non-physical displacement to each particle at every time step, moving it towards regions of lower particle concentration. This process helps to maintain a more homogeneous particle distribution throughout the simulation, which significantly enhances the long-term stability and accuracy of the SPH method.
\end{itemize}

\section{Programming Framework}
A robust and extensible software architecture is the cornerstone of successful numerical simulation software. This section delves into the design architecture and the functionality of key modules within FEALPy.CFD, demonstrating how it achieves an organic unification of high-performance computing and flexible application development. 
This chapter is divided into three main parts. 

\subsection{The Design Architecture of FEALPy}
\label{sec2} 
\begin{figure}[!htbp]
	\centering
	\includegraphics[width=0.85\textwidth]{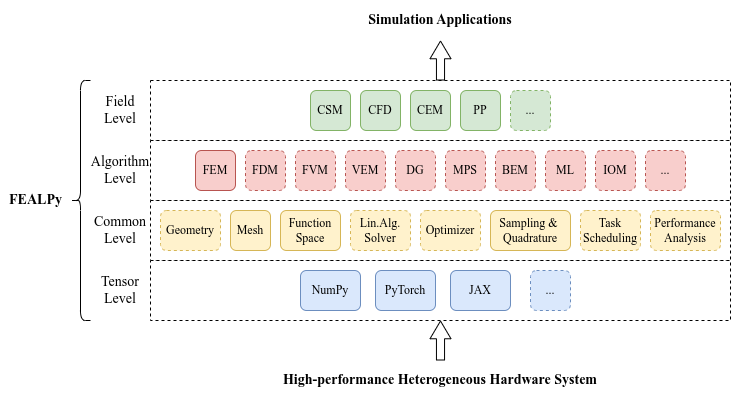}
	\caption{The leveled architecture of FEALPy, comprising tensor, common, algorithm, and field levels, progressing from low-level functionalities to high-level applications.}
	\label{fealpy_architect}
\end{figure}

As a Python library for CAX simulation and computing, FEALPy adopts a leveled modular architecture, as illustrated in Figure ~\ref{fealpy_architect}. This architecture is structured from bottom to top into four distinct levels: the Tensor Level, the Common Level, the Algorithm Level, and the Domain Level. Each level encapsulates specific functional modules, which collectively form an extensible and high-performance simulation platform. The interface currently adheres to the Python array API standard v2023.12 \cite{pythonAPI}, ensuring compatibility with multiple backends such as NumPy, PyTorch, and JAX. This design achieves the goal of adapting the same codebase to different software and hardware platforms.

The functions of each level provide support for FEALPy.CFD in various aspects, as detailed below:
\begin{itemize}
	\item \textbf{Tensor Level}:Serving as the cornerstone of the entire framework, this level provides a unified tensor operation interface and abstracts away the implementation differences of various backends like NumPy, PyTorch, and JAX through a backend manager. This hardware and computation abstraction mechanism is the core foundation that enables FEALPy.CFD to achieve cross-platform, high-performance computing.
	\item \textbf{Commonality Level}：This level provides a comprehensive set of fundamental numerical infrastructure for FEALPy.CFD. It not only includes flexible mesh data structures and rapid generation tools for simple geometric domains but also encapsulates core functionalities essential for building Finite Element Methods, such as various function spaces and the computation of basis functions and their derivatives. This greatly accelerates the development efficiency of pre-processing and discretization in CFD applications.
	\item \textbf{Algorithm Level}:This level encapsulates a variety of core numerical algorithm components, including general-purpose linear algebra solvers and essential kernels for FEM and FVM, such as numerical integration and matrix assembly. This provides FEALPy.CFD with validated and directly callable modules, allowing it to focus on high-level application logic and solution algorithms while avoiding the redundant development of low-level algorithms.
	\item \textbf{Field Level}:This level is the part of the FEALPy architecture directly oriented towards specific physics applications. It enables FEALPy.CFD to be coupled with other field modules for co-simulation, providing architectural-level support for tackling multi-physics and interdisciplinary challenges.
\end{itemize}
In summary, FEALPy's leveled architecture achieves a high degree of functional decoupling across its different levels. This approach simultaneously ensures functional completeness and extensibility while maintaining high performance and cross-platform portability. Consequently, it allows developers using FEALPy.CFD to focus on high-level algorithms or application logic without needing to be concerned with the differences in underlying hardware or libraries.

\subsection{The Design Architecture of FEALPy.CFD}
As part of the Field Level at the top of the FEALPy architecture, FEALPy.CFD is a simulation library specifically designed for computational incompressible fluid dynamics applications. It fully leverages the solid foundation provided by the lower layers of FEALPy, inheriting its unified backend management for tensor computations, rich collection of numerical algorithm components, and versatile mesh and geometry processing capabilities. Building upon this base, FEALPy.CFD further establishes a highly modular upper-level application architecture tailored to the specific characteristics of fluid problems.This "inherit and extend" design pattern enables FEALPy.CFD to not only full leverage the advantages of the underlying FEALPy library but also, through its own high-level architecture optimized for fluid dynamics, to significantly enhance the usability, flexibility, and computational efficiency for solving complex flow problems.

\begin{figure}[!htbp]
	\centering
	\includegraphics[width=0.9\textwidth]{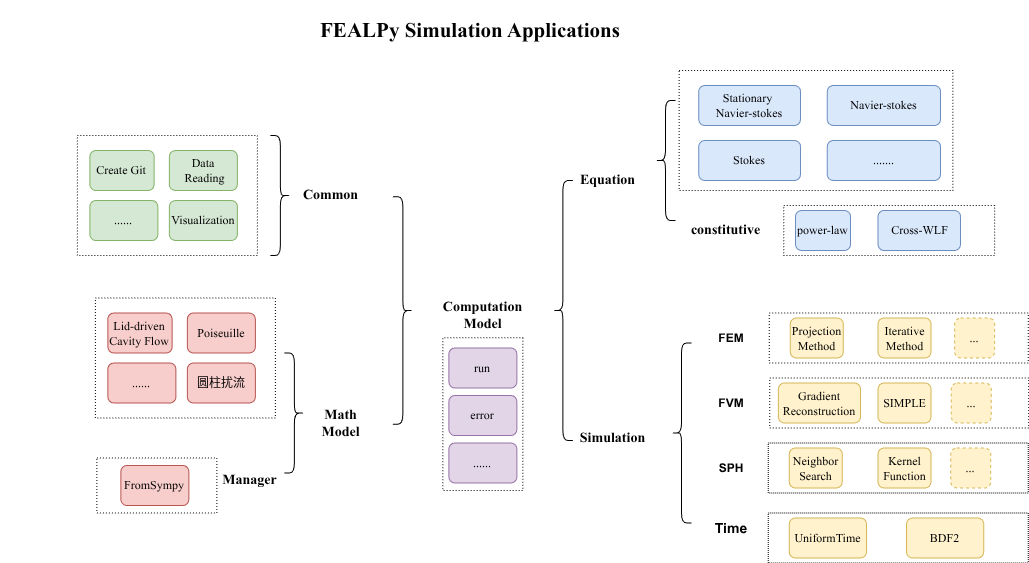}
	\caption{The core architecture of FEALPy.CFD, illustrating its design around the four central modules of MathModel, Equation, Simulation, and Common, aimed at separating the physical problem description, numerical methods, and case settings.}
	\label{cfd_architect}
\end{figure}

As illustrated in the Figure ~\ref{cfd_architect}, the core architecture of FEALPy.CFD is constructed around four key modules: the MathModel module, the Equation module, the Simulation module, and the Common module. This design clearly separates the physical description of the problem (Equation), the numerical solution method (Simulation), and the specific example settings (MathModel). Under the central management of the Computation Model, these modules collaborate through well-defined interfaces to form a unified computational framework that supports multiple physics equations and numerical methods. This architecture endows FEALPy.CFD with outstanding extensibility and flexibility, enabling it to conveniently address diverse simulation challenges in incompressible fluid dynamics.

\subsubsection{The Equation Module}
In the FEALPy.CFD framework, the Equation module serves as the physical modeling component. Its primary responsibility is to convert the mathematical model descriptions defined in the MathModel into structured computational objects that can be called by the Simulation module. The design of this module incorporates several key functionalitie:

\begin{itemize}
	\item \textbf{Flexible Physical Modeling}:
\end{itemize}
 \qquad This module not only manages the physical coefficients of various terms in the equations (such as viscosity and inertial terms) and supports their variation with time or space (e.g., in multiphase flow models), but it also has built-in support for multiple constitutive equations. Users can easily switch between the standard Newtonian fluid model and complex non-Newtonian fluid models (e.g., the Power-law or Cross-WLF models), which greatly expands the application scope of the framework.
 \begin{itemize}
 	\item \textbf{Boundary Condition Processing}: 
 \end{itemize}
\qquad This module is responsible for parsing the types of boundary conditions (e.g., Dirichlet, Neumann) defined in the MathModel object. It provides a clear interface for the Simulation to correctly apply these boundary constraints.
\begin{itemize}
	\item \textbf{Parameter Management}:
\end{itemize}
\qquad To enhance user experience and debugging efficiency, the Equation module offers a convenient parameter management function. Users can intuitively inspect the current status of all physical parameters with a simple print command, which facilitates visualization and significantly improves the transparency and reliability of the model setup.

\subsubsection{The Common Module}
The Common module is a public functional component within the FEALPy.CFD framework, the design of this module adheres to the principle of algorithm agnosticism, centralizing and encapsulating common tasks in the simulation workflow that are independent of specific numerical methods. This approach enhances code maintainability and the overall integrity of the framework.

The core functionalities of this module primarily cover two aspects:

\begin{itemize}
	\item \textbf{Data Input/Output and Format Compatibility}: 
\end{itemize}
\qquad The module provides read/write interfaces for standard scientific computing visualization format files (such as VTK's .vtu files). This feature not only facilitates the import of externally generated meshes or initial field data into the solver but also enables the computational results from FEALPy.CFD to be seamlessly integrated with mainstream post-processing software like ParaView for in-depth analysis. This enhances the openness and compatibility of the framework.
\begin{itemize}
		\item \textbf{Result Visualization and Presentation}:
\end{itemize}
 \qquad For immediate feedback and rapid validation, the module integrates quick plotting functionalities based on Matplotlib, allowing users to instantly generate 2D contour plots,convergence rate plots, or vector field plots during or after the computation. Furthermore, for transient problems, the module supports the convenient rendering of time-series results into animated images in GIF format. This functionality is particularly important for quickly evaluating computational results, creating presentation materials, and intuitively demonstrating dynamic processes in academic reports, thereby greatly improving user convenience.

\subsubsection{The Simulation Module}
The Simulation module is the computational core of the FEALPy.CFD framework. It is responsible for the numerical discretization of the physical model defined in the Equation module, encompassing both time and spatial discretization. Additionally, it integrates various numerical solution algorithms for different discretization methods.  Its core features are as follows:

\begin{itemize}
	\item \textbf{Support for Multiple Discretization Methods}:
\end{itemize}
\qquad A significant advantage of this module is its native support for several mainstream spatial discretization methods. It currently has built-in support for three major categories: FEM, FVM, and SPH. Concurrently, it features an independent time discretization module to manage various time-stepping schemes, such as Backward Differentiation Formulas(BDF).

Integrating these methods within a unified framework offers a dual benefit: On one hand, users can select the optimal numerical method for a specific problem to achieve the best computational efficiency and accuracy. On the other hand, it provides a fair comparison platform for algorithm researchers to evaluate the performance of different methods within the same data structures and environment. This also establishes FEALPy.CFD as a comprehensive computational platform capable of tackling different types of flow problems and facilitating multi-method comparisons.
\begin{itemize}
	\item \textbf{Object-Oriented Modular Design}:
\end{itemize}
\qquad Within each discretization method, the Simulation module is designed following an object-oriented programming paradigm. Each method (e.g., FEM) provides a base class responsible for managing generic operations pertinent to it, such as defining function spaces, selecting numerical quadrature rules, and managing degrees of freedom. Specific numerical algorithms—for instance, various projection methods for the NS equations or different iterative schemes are implemented as derived classes.

This design greatly enhances code reusability and makes the process of adding new numerical algorithms exceptionally clear and convenient.
\begin{itemize}	
	\item \textbf{Process Decomposition and Componentization Design Philosophy}:
\end{itemize}
\qquad The core design philosophy of the Simulation module is not to provide a monolithic, fixed solution process, but rather to adopt a strategy of process decomposition and componentization. Taking the classic IPCS pressure-correction algorithm as an example, the module does not offer a single, all-encompassing solve function. Instead, it breaks down the entire algorithm into multiple independent steps, such as momentum prediction, the pressure Poisson equation, and velocity correction. More critically, it provides independent and feature-rich programming interfaces for each step, such as interfaces for assembling matrices (BForm) and right-hand side vectors (LForm), as well as for updating the coefficients of different terms in each step (update).

This "divide and conquer" design approach yields two main advantages: First, the high-level ComputationModel can act as a "workflow orchestrator", conveniently calling these standardized step-components to integrate them into a stable and reliable complete solver in a predefined sequence. Second, it offers high flexibility for research and development: algorithm researchers can interact directly with these fine-grained step interfaces. They can implement improvements to existing algorithms or develop prototypes for new ones by replacing, modifying, or reorganizing any of these stages , thereby gaining maximum freedom and extensibility.

\subsubsection{The MathModel Module}
The MathModel module is the problem definition center of FEALPy.CFD. It provides structured and reusable mathematical model descriptions for both the user and the framework's Equation module. This module not only includes a rich library of benchmark cases but also offers powerful support for users to define new, custom problems.

Its core functions and features include:

\begin{itemize}
	\item \textbf{Built-in Benchmark Case Library}:
\end{itemize}
\qquad The module contains a set of classic test problems from the CFD field. This includes verification and validation cases from the literature that have exact analytical solutions (covering both 2D and 3D), which are used to accurately test the convergence order of numerical methods. It also covers standard benchmark problems, such as pipe flow and flow past a cylinder, which are used to validate the effectiveness of numerical algorithms by comparing computational results with published numerical or experimental data.
\begin{itemize}
	\item \textbf{Convenient Interfaces for Custom Problems}:
\end{itemize}

To maximize the framework's flexibility and extensibility, the MathModel module also provides users with two ways to define custom problems:

From Sympy: This is a convenient rapid-testing tool that allows users to define arbitrary analytical solutions using the Sympy symbolic computation library \cite{Sympy}. The framework automatically handles tedious steps like differentiation and substitution to generate the corresponding source terms and boundary conditions, greatly simplifying the verification process for new algorithms.

Standardized PDE Interface: The module defines a clear and standardized PDE programming interface. By adhering to this interface, users can encapsulate their unique physical problems (such as new geometric domains, boundary conditions, or source terms) into PDE objects that are fully compatible with the built-in cases. These objects can then be seamlessly passed to the Equation module for subsequent physical model parsing.

The design of the Model module makes FEALPy.CFD both an out-of-the-box solution tool with a rich set of examples and a highly extensible research platform that supports secondary development by users.

\subsection{Module Collaboration}
The power and flexibility of the FEALPy.CFD from its distinct modules, each  performing its specific role while collaborating closely. In the design of FEALPy.CFD, the ComputationModel module acts as the workflow orchestrator, responsible for integrating and driving the other modules to form a complete, linear simulation workflow. As illustrated in the figure ~Figure \ref{collaboration}, a typical solution process follows these steps:

\begin{figure}[!htbp]
	\centering
	\includegraphics[width=0.95\textwidth]{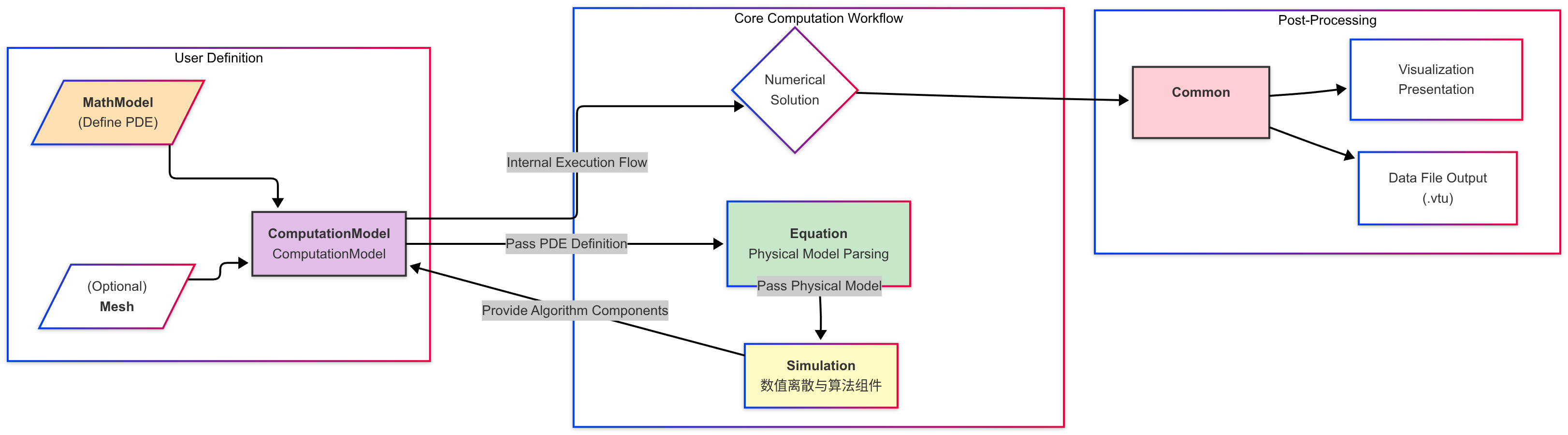}
	\caption{Architectural flowchart of the FEALPy.CFD solver. }
	\label{collaboration}
\end{figure}

\begin{enumerate}
	\item \textbf{Problem Definition}: The workflow begins with the user's definition. The MathModel module is responsible for defining or loading a specific PDE problem, including its geometric domain, boundary conditions, and source terms, as well as various initial mesh types (e.g., quadrilateral, triangular, tetrahedral, polygonal). Users can also pass in their own custom-constructed Mesh objects as needed.
	\item \textbf{Physical Model Parsing}: The PDE object defined in the MathModel is passed to the Equation module. This module performs a physical-level parsing, identifying the various components of the equation and their corresponding physical coefficients.
	\item \textbf{Numerical Discretization}: The parsed physical model is then handed over to the Simulation module for processing. This module is the core of numerical discretization, where the user can select specific spatial discretization schemes (FEM, FVM, SPH) and the corresponding numerical solution algorithms (e.g., Newton's method, SIMPLE algorithm).
	\item \textbf{Execution and Solution}: Finally, the ComputationModel module acts as the main controller. It calls the low-level algorithm components provided by the Simulation module (such as interfaces for matrix and vector assembly) and integrates functionalities like the selection of linear solvers (e.g., GMRES, MINRES, spsolve) and error calculation to execute the complete solution process.
	\item \textbf{Post-processing and Visualization}: After the solution is complete, the computational results are passed to the Common module for post-processing. This module offers extensive visualization support. Users can either leverage its built-in interfaces to call libraries like Matplotlib for quick 2D result presentation or export the data into standard VTK format files for in-depth 3D visualization and analysis in professional post-processing software like ParaView.
\end{enumerate}

\subsection{Multi-Backend Switching}
To effectively address the diverse computational resource requirements for problems of different scales, from rapid prototyping to production-level industrial simulations. FEALPy.CFD inherits and fully leverages the hardware abstraction capabilities of FEALPy's underlying Tensor Level. This enables a flexible multi-backend support architecture. Users can freely switch between multiple tensor computation backends such as NumPy, PyTorch, and JAX to adapt to various software and hardware platforms, the advantages and applicable scenarios for different backends are as follows:

\textbf{NumPy Backend}: Based on CPU computation, NumPy is the cornerstone of scientific computing in Python. Its user-friendly interface and ease of debugging make it highly suitable for rapid prototyping, algorithm validation, and small to medium scale computational tasks.

\textbf{JAX Backend}: Leveraging its Just-In-Time (JIT) compilation feature, JAX can compile Python code into highly optimized machine code, achieving significant performance gains on both CPUs and GPUs. Its built-in automatic differentiation capabilities also open up possibilities for future integration with methods like Physics-Informed Neural Networks (PINNs).

\textbf{PyTorch Backend}: As a leading framework in the deep learning field, PyTorch boasts a powerful ecosystem and mature GPU support, making it an ideal choice for performing large-scale, computationally intensive simulations. It also facilitates the deep integration of traditional CFD simulations with artificial intelligence models.

The most crucial advantage of this architecture is that the high-level API of FEALPy.CFD remains consistent, regardless of the underlying computation backend selected. This design completely decouples algorithm implementation from hardware execution. As a result, users can migrate and accelerate their work across different platforms without modifying the main body of their code, which dramatically improves user productivity and software portability.

\section{Getting Started with FEALPy.CFD}
To intuitively demonstrate the design philosophy and the concise, efficient workflow of the FEALPy.CFD framework, and to lower the learning curve for potential users, this chapter provides a detailed, from-scratch usage tutorial. Our goal is to guide users—especially beginners in the CFD field or researchers seeking efficient development tools—to master the core functionalities of this framework in the shortest possible time. This chapter is divided into two main parts: first, we will detail the installation steps for FEALPy to ensure readers can correctly configure their development environment; subsequently, we will demonstrate the workflow of FEALPy.CFD through a pedagogical case study involving an equation with a prescribed analytical solution.

\subsection{Software Installation}
FEALPy.CFD is a functional module of the FEALPy finite element library. Therefore, users do not need to install FEALPy.CFD independently; usage is enabled simply by completing the installation of the main FEALPy library

Core installation steps:
\begin{enumerate} 
	\item Clone the FEALPy repository from GitHub:
	\begin{lstlisting}[language=bash]
	git clone https://github.com/weihuayi/fealpy.git
	\end{lstlisting}
	\item Change into the FEALPy directory and install it in editable mode:
	\begin{lstlisting}[language=bash]
	cd fealpy
	pip install -e . 
	\end{lstlisting}
	\item Run the following command in the Python interpreter to verify the FEALPy installation.
	\begin{lstlisting}[language=bash]
	import fealpy 
	print(fealpy.__version__) 
	\end{lstlisting}
\end{enumerate}

\subsection{Example:The Steady-State Navier-Stokes Equations}
Following the environment configuration, this section demonstrates the complete computational workflow in FEALPy.CFD through a concrete numerical example. We consider a problem for the steady-state incompressible NS equations \eqref{steady_ns_equation}, defined on the computational domain $\Omega = [0,1]^2$, which possesses a known analytical solution 

\begin{equation}
	\begin{cases}
		u=10x^2(x-1)^2y(y-1)(2y-1), \\
		v=-10x(x-1)(2x-1)y^2(y-1)^2, \\
		p=10(2x-1)(2y-1). & 
\end{cases}
\end{equation}
We can derive the corresponding source term and  boundary conditions by substituting this exact solution into the governing equations.

The first step is to import the required modules. This includes core modules from the FEALPy library, such as the computational backend, as well as the necessary components from FEALPy.CFD: the solver Model, the PDE selection module, and the visualization module.

\begin{lstlisting}[language=python, label={code:import}]
	from fealpy.backend import backend_manager as bm
	from fealpy.cfd.model import CFDTestModelManager
	from fealpy.cfd.common.visualize import output, showmultirate
	from fealpy.cfd import StationaryIncompressibleNSLFEMModel
\end{lstlisting}

We define the number of mesh refinements, set up the initial grid, and then instantiate the appropriate solver Model.

\begin{lstlisting}[language=python, label={code:mesh}]
	maxit = 5
	manager = CFDTestModelManager('stationary_incompressible_navier_stokes')
	pde = manager.get_example(3)
	mesh = pde.init_mesh['uniform_tri'](nx=4, ny=4)
	model = StationaryIncompressibleNSLFEMModel(pde, mesh)
\end{lstlisting}

 We select the solution method, define the finite element function spaces, and generate data logs for subsequent error analysis and visualization. These can be customized by the user based on their requirements. In this implementation, the velocity field is defined in a continuous, second-order Lagrange space, while the pressure field is in a continuous, first-ordeLagrange space
\begin{lstlisting}[language=python, label={code:method}]
	model.method['Newton']()   
	fem = model.fem
	fem.set.uspace('Lagrange', p=2)
	fem.set.pspace('Lagrange', p=1)
	errorMatrix = bm.zeros((2, maxit), dtype=bm.float64)
	NDof = bm.zeros((2, maxit), dtype=bm.float64)
\end{lstlisting}

 Next executes the main solution routine. To control the stopping criteria for the iterative process, the Model is configured with a maximum iteration limit of 10 and a convergence tolerance of 
\begin{lstlisting}[language=python, label={code:simulation}]
	for i in range(maxit):
		uh, ph = model.run(maxstep=10, tol=1e-10)
		u_errorMatrix[0, i] = mesh.error(uh, pde.velocity)
		p_errorMatrix[0, i] = mesh.error(ph, pde.pressure)
		if i < maxit - 1:
		mesh.uniform_refine()
		fem.update_mesh(mesh)
\end{lstlisting}

Finally, the previously stored error data is analyzed to generate a plot of the convergence rates. The results are then output for visualization

\begin{lstlisting}[language=python, label={Post-processing}]
from matplotlib import pyplot as plt
errorType = ['$|| u - u_h||_{L2}$ ','$|| p - p_h||_{L2}$ ']
showmultirate(plt, 0, NDof, errorMatrix,  errorType, propsize=20)
plt.show()
output(mesh, {'velocity':uh, 'pressure':ph}, 
name='result.vtu',path = './')
\end{lstlisting}

\begin{figure}[!htbp]
	\centering
	\subfloat[Pressure]{
		\includegraphics[width=0.45\textwidth]{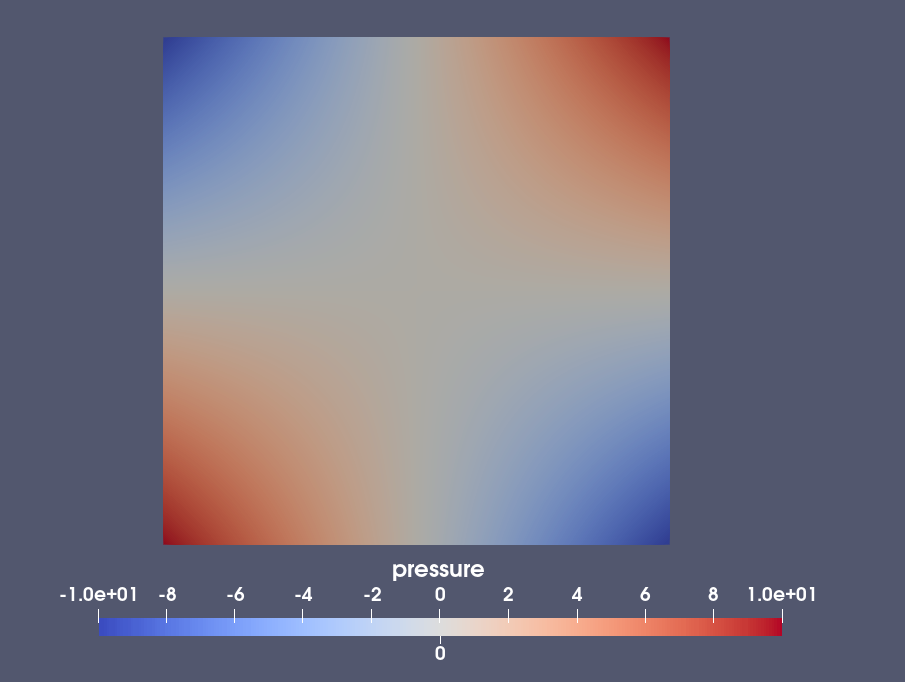}
		\label{fig:Getting Started pressure}
	}
	\subfloat[Velocity Magnitude]{
		\includegraphics[width=0.44\textwidth]{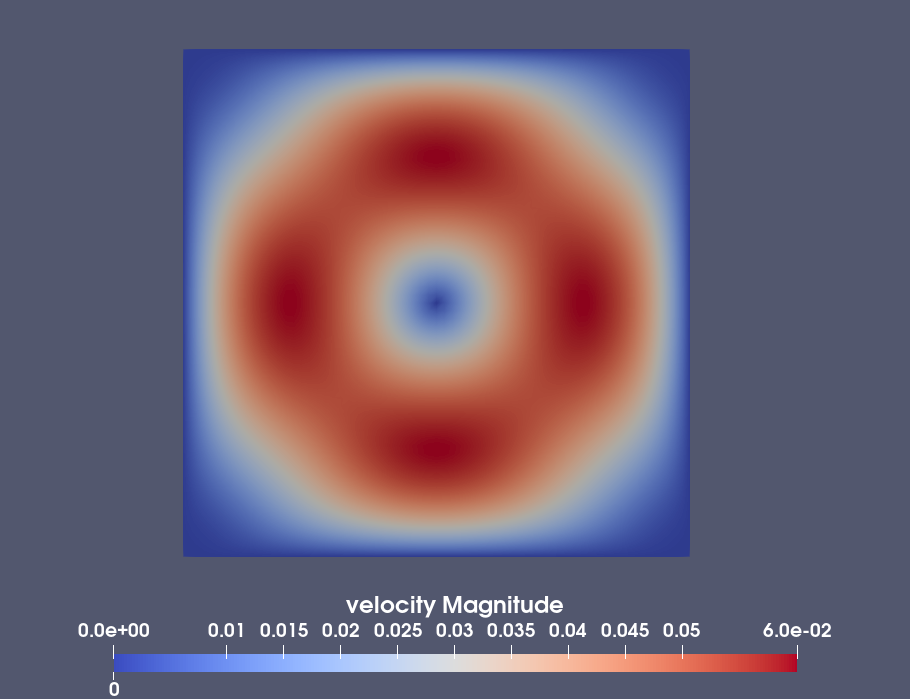}
		\label{fig:Getting Started velcotity}
	}
	\caption{The result of steady-state incompressible NS equations}
	\label{fig:Getting Started}
\end{figure}

\begin{figure}[!htbp]
	\centering
	\includegraphics[width=0.8\textwidth]{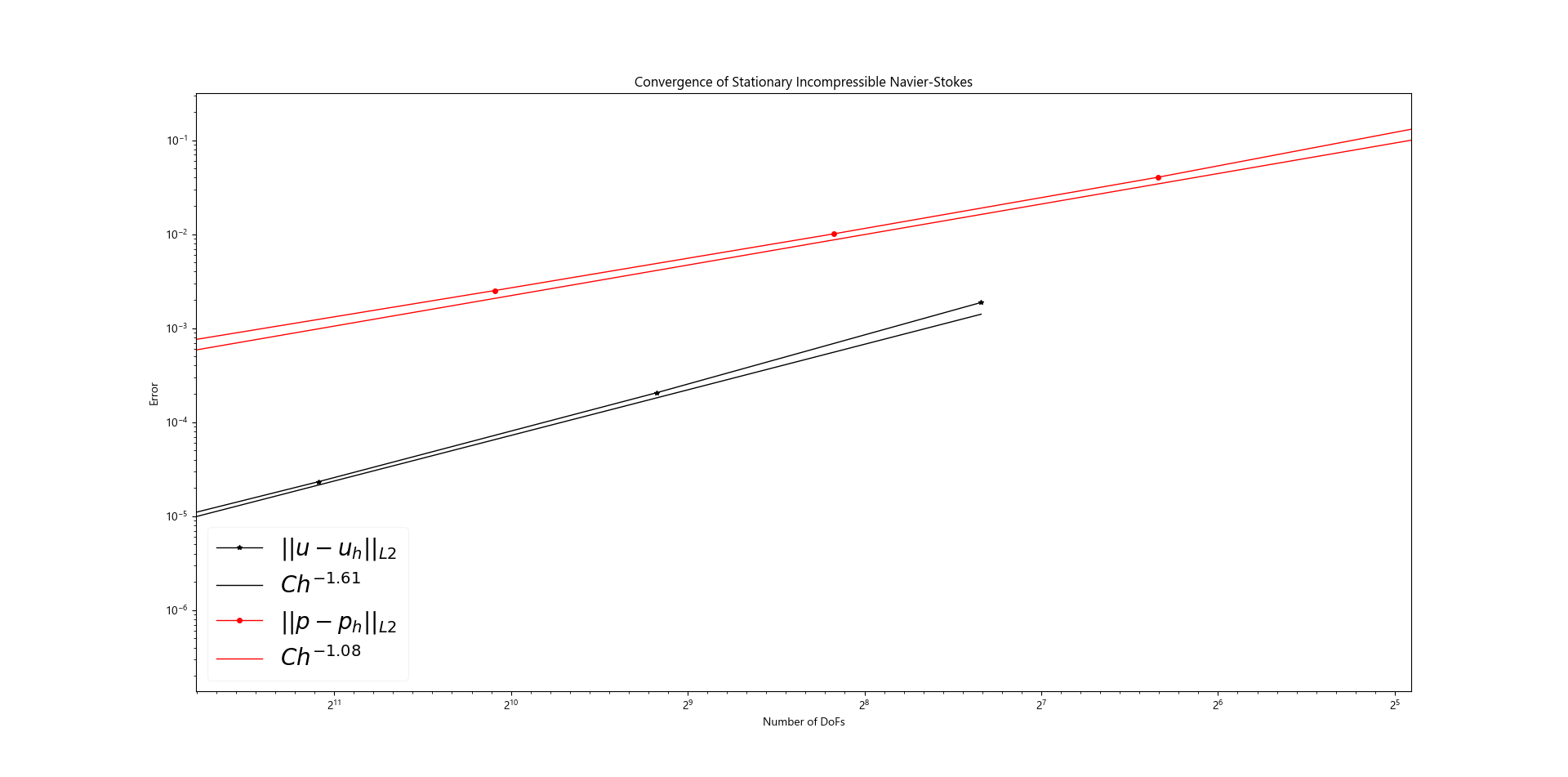}
	\caption{The convergence of steady-state incompressible NS equations}
	\label{fig:convergence of steady-state}
\end{figure}

The results presented in Table \ref{tab:convergence_data} , Figures \ref{fig:Getting Started} and \ref{fig:convergence of steady-state} validate the numerical accuracy and theoretical convergence order of the implemented steady-state solver.

\begin{table}[htbp]
	\centering
	\caption{Convergence analysis for velocity and pressure}
	\label{tab:convergence_data}
	\begin{tabular}{||r|c|c|r|c|c||} 
		\hline
		% --- 表头 ---
		$u_{\text{Dofs}}$ & $\| \mathbf{u} - \mathbf{u}_h \|_{L_2}$ & k & $p_{\text{Dofs}}$ & $\| p - p_h \|$ & k \\
		\hline
		% --- 数据行 ---
		162  & 1.87952236e-03 &              & 25   & 0.1625292  &              \\
		578  & 2.05935706e-04 & 3.19010018   & 81   & 0.04039637 & 2.00840133   \\
		2178 & 2.32700663e-05 & 3.14564677   & 289  & 0.01008773 & 2.00162386   \\
		8450 & 2.80719074e-06 & 3.05127632   & 1089 & 0.00252153 & 2.00022969   \\
		33282& 3.47550861e-07 & 3.01383107   & 4225 & 0.00063037 & 2.00002911   \\
		\hline
	\end{tabular}
\end{table}

\section{Numerical Examples}
To comprehensively validate the accuracy, efficiency, and architectural flexibility of the FEALPy.CFD solver, as well as to demonstrate its potential in solving complex flow and multiphysics problems, this chapter presents a series of carefully designed and representative numerical examples. The organization of these cases follows a logical progression of increasing complexity, moving from fundamental verification to the exploration of advanced applications.

\subsection{Verification of Framework Flexibility and Numerical Accuracy}
This part is dedicated to the verification of the FEALPy.CFD solver's various functionalities, with a focus on quantitatively demonstrating its numerical accuracy and framework flexibility. Using the code from the tutorial chapter as a baseline, we will illustrate how the framework's API design facilitates simple and intuitive switching between different meshes, equations, orders of spatial accuracy, and numerical methods. The collective results of these tests provide solid evidence of the framework's reliability, flexibility, and numerical precision.

All tests in this part are based on a classical benchmark problem for the unsteady NS equations \eqref{ns_equation}, for which the exact solution is given by \cite{JOHN}:
\begin{equation}
	\begin{cases}
		u_1(x,y)=2\pi\sin(t)\sin^2(\pi x)\sin(\pi y)\cos(\pi y), \\
		u_2(x,y)=-2\pi\sin(t)\sin(\pi x)\cos(\pi x)\sin^2(\pi y), \\
		p(x,y)=20\sin(t)(x^2y-\frac{1}{6}) & 
	\end{cases}
\end{equation}
The problem is defined on the unit square domain $\Omega = (0, 1)\times(0,1)$ with a viscosity  $\nu = 1$ and a density $\rho = 1$. As a benchmark case with a known analytical solution, it is ideally suited for verifying the numerical orders of convergence.

To switch from a steady-state to an unsteady NS solver, the ComputationModel must be changed accordingly. This test case can be conveniently loaded via the CFDTestModelManager.

\begin{lstlisting}[language=python,label={code:SC}]
from fealpy.cfd import StationaryIncompressibleNSLFEMModel
pde = manager.get_example(3)
model = StationaryIncompressibleNSLFEMModel(pde=pde)
\end{lstlisting}

\begin{figure}[!htbp]
	\centering
	\subfloat[uniform quadrilateral mesh]{
		\includegraphics[width=0.35\textwidth]{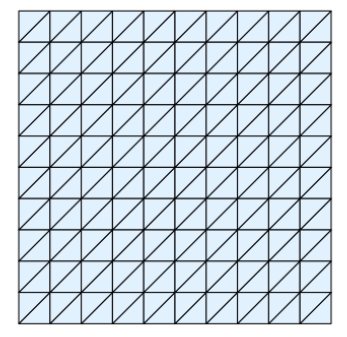}
		\label{fig:tri_mesh}
	}
	\subfloat[uniform triangular mesh]{
		\includegraphics[width=0.35\textwidth]{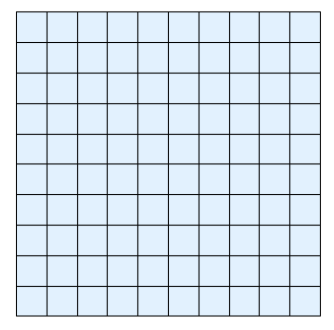}
		\label{fig:quad_mesh}
	}
	\caption{Different Mesh}
	\label{fig:Switch Mesh}
\end{figure}

We begin by verifying the solver's accuracy across different mesh. For this purpose, a baseline configuration employing Newton's method and the P2-P1 Taylor-Hood element is used to perform convergence tests on uniform triangular and quadrilateral meshes. Users can easily switch the mesh type by simply modifying the init\_mesh parameter in the pde object:

\begin{lstlisting}[language=python, label={code:SM}]
mesh = pde.init_mesh['uniform_quad'](nx=10, ny=10)
\end{lstlisting}

The different mesh topologies are illustrated in Figure \ref{fig:Switch Mesh}.The error convergence results for the different mesh topologies are summarized in Table \ref{tab:convergence_different_mesh}. 

\begin{table}[htbp]
	\centering
	\caption{Convergence analysis for two different mesh}
	\label{tab:convergence_different_mesh}
	
	% --- 左边表格 ---
	\begin{minipage}[t]{0.45\textwidth}
		\centering
		\caption*{uniform quadrilateral mesh}
		\begin{tabular}{||c|c|c|c|c||}
			\hline
			$1/h$ & $\| \mathbf{u} - \mathbf{u}_h \|_{L_2}$ & k & $\| p - p_h \|_{L_2}$ & k \\
			\hline
			4  & 3.226e-02 & - & 2.229e-01 & - \\
			8  & 4.307e-03 & 2.90 & 1.724e-02 & 3.69 \\
			16 & 5.660e-04 & 2.93 & 2.994e-03 & 2.52 \\
			32 & 7.142e-05 & 2.99 & 7.317e-04 & 2.03 \\
			\hline
		\end{tabular}
	\end{minipage}
	\hfill
	% --- 右边表格 ---
	\begin{minipage}[t]{0.45\textwidth}
		\centering
		\caption*{uniform triangular mesh}
		\begin{tabular}{||c|c|c|c|c||}
			\hline
			$1/h$ & $\| \mathbf{u} - \mathbf{u}_h \|_{L_2}$ & k & $\| p - p_h \|_{L_2}$ & k \\
			\hline
			4  & 7.739e-02 & - & 1.776e+00 & - \\
			8  & 8.607e-03 & 3.17 & 1.505e-01 & 3.56 \\
			16 & 1.034e-03 & 3.06 & 1.219e-02 & 3.63 \\
			32 & 1.300e-04 & 3.00 & 1.835e-03 & 2.73 \\
			\hline
		\end{tabular}
	\end{minipage}
\end{table}

To further showcase the framework's flexibility in accommodating different solution algorithms,we switch from the fully coupled Newton's method to the decoupled Incremental IPCS。Highlighting the "plug-and-play" nature of the architecture, this algorithmic switch is achieved with a single line of code:

\begin{lstlisting}[language=python,label={code:SM}]
model.method['IPCS']()        
\end{lstlisting}

The error convergence results for the IPCS algorithm are summarized in Table \ref{tab:convergence_ipcs}.

\begin{table}[htbp]
	\centering
	\caption{IPCS convergence analysis}
	\label{tab:convergence_ipcs}
	
	\begin{tabular}{||c|c|c|c|c||}
		\hline
		$1/h$ & $\| \mathbf{u} - \mathbf{u}_h \|_{L_2}$ & k & $\| p - p_h \|_{L_2}$ & k \\
		\hline
		4  & 0.018602019897807 & - & 0.16210974061646600 & - \\
		8  & 0.0024489585393050 & 2.93 & 0.01963367604120530 & 3.05 \\
		16 & 0.000310221142660 & 2.98 & 0.003667827451801 & 2.42 \\
		32 & 0.000039010906084 & 2.99 & 0.00087284139087142 & 2.07 \\
		\hline
	\end{tabular}
\end{table}

We then validate the framework's support for higher-order finite element spaces. To this end, the approximation is elevated by employing the P3-P2 Taylor-Hood element, which uses third-order polynomials for the velocity space and second-order polynomials for the pressure space. The system is solved using the IPCS method. As before, the required code modifications are highly intuitive:

\begin{lstlisting}[language=python,label={code:SM}]
fem.set.uspace('Lagrange', p=3)
fem.set.pspace('Lagrange', p=2)    
\end{lstlisting}
The error convergence results for the P3-p2 finite element spaces are summarized in Table \ref{tab:convergence_p3_p2}.

\begin{table}[htbp]
	\centering
	\caption{P3-P2 Convergence analysis}
	\label{tab:convergence_p3_p2}
	
	\begin{tabular}{||c|c|c|c|c||}
		\hline
		$1/h$ & $\| \mathbf{u} - \mathbf{u}_h \|$ & k & $\| p - p_h \|$ & k \\
		\hline
		4  & 0.01418658541424 & - & 0.158195879127172 & - \\
		8  & 0.000857410621868 & 4.05 & 0.02094076704491451 & 2.91 \\
		16 & 5.37655289798662e-05 & 4.00 & 0.00250455020432984 & 3.06 \\
		32 & 3.341004258893174e-06 & 4.01 & 0.0002933563079698 & 3.09 \\
		\hline
	\end{tabular}
\end{table}

Finally, to verify the FEALPy.CFD framework's three-dimensional (3D) simulation capabilities, we select the Poiseuille flow in a square duct as a benchmark case. This canonical problem in fluid mechanics simulates the laminar flow of a fluid through a long, straight duct with a square cross-section, driven by a constant pressure gradient.

\begin{figure}[!htbp]
	\centering
	\includegraphics[width=0.65\textwidth]{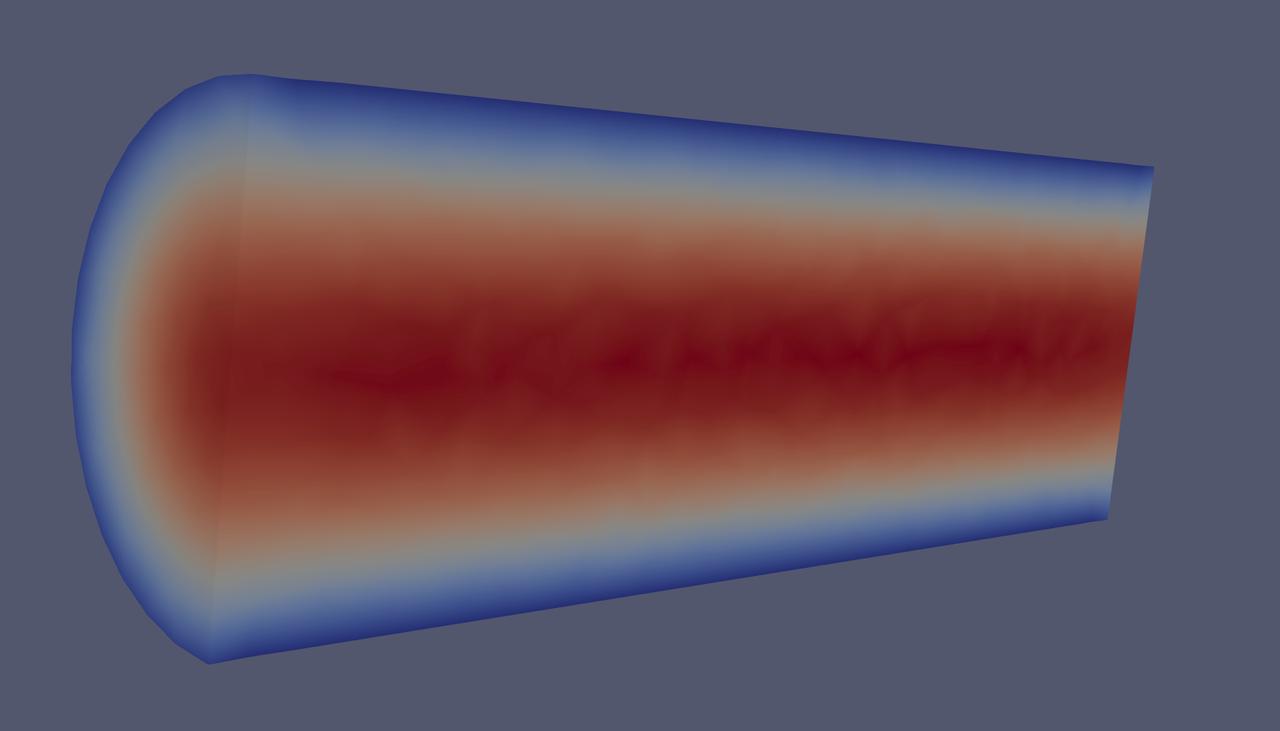}
	\caption{3D Poiseuille Flow}
	\label{Poiseuille flow}
\end{figure}

The Figure \ref{Poiseuille flow} is the result of 3D Poiseuille flow , presents the steady-state velocity profile in the square duct, as computed by FEALPy.CFD. The classic paraboloid shape is clearly visible, where the velocity peaks at the duct's centerline and smoothly decays to zero at the walls, satisfying the no-slip boundary condition. These results are in excellent qualitative agreement with the theoretical solution for Poiseuille flow, providing a clear visual verification that the solver accurately captures the fundamental characteristics of three-dimensional internal flows.

\subsection{Flow past a cylinder}
The objective of this section is to validate the accuracy and robustness of the FEALPy.CFD solver for capturing complex flow phenomena, particularly periodic vortex shedding, by using a classic computational fluid dynamics benchmark. We first selected the 2D flow in a channel around a fixed cylinder as proposed by Schäfer and Turek \cite{TUREK}, and specifically address the 2D-2 test case from the featflow DFG benchmark suite \cite{featflow}.Following this validation, we further present the simulation of flow past a standard airfoil (NACA 0012) to demonstrate FEALPy.CFD's powerful geometric handling capabilities and its numerical robustness on more complex, arbitrary geometries.

\begin{figure}[!htbp]
	\centering
	\subfloat[Pressure]{
		\includegraphics[width=0.7\textwidth]{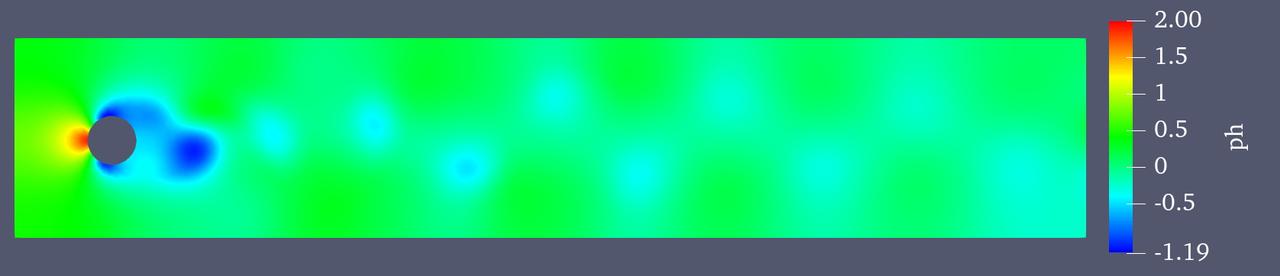}
		\label{}
	}\\
	\subfloat[Streamlines]{
		\includegraphics[width=0.7\textwidth]{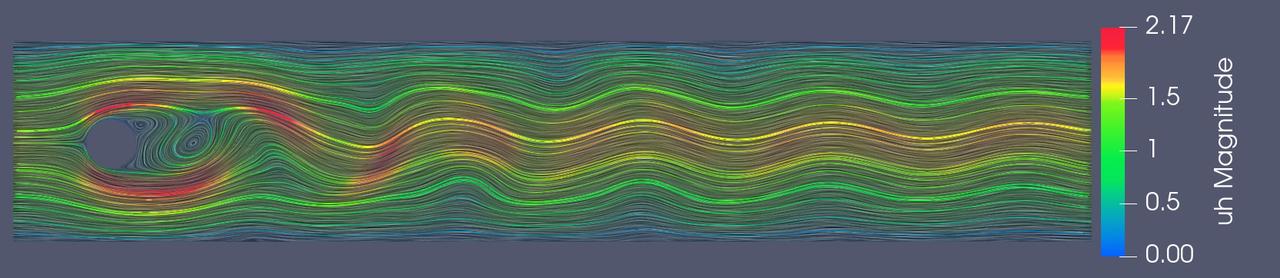}
		\label{} 
	}\\
	\subfloat[Velocity Magnitude]{
		\includegraphics[width=0.7\textwidth]{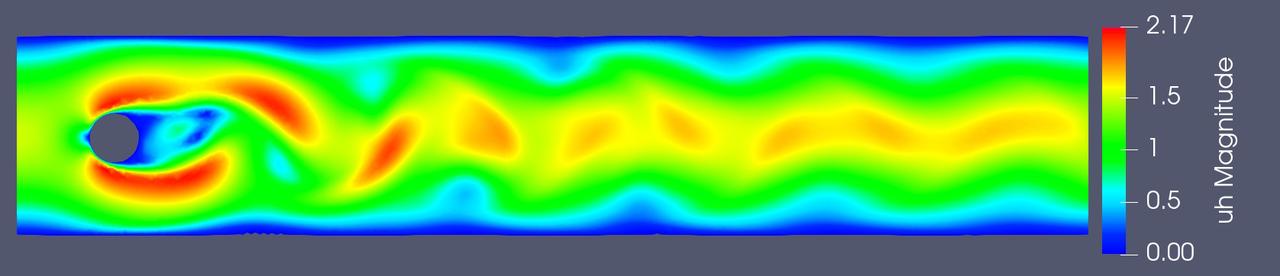}
		\label{}
	}
	\caption{The result of Flow past a cylinder}
	\label{fig:flow_past_cylinder}
\end{figure}

Figure \ref{fig:flow_past_cylinder} provides a qualitative comparison of the instantaneous flow fields, including velocity magnitude, pressure distribution, and streamlines, comparing our results with the reference result from the featflow \cite{featflow}. The simulation results from FEALPy.CFD successfully reproduce the key topological features of the flow, most notably the periodic Kármán vortex street in the cylinder's wake. The overall morphology, size, and shedding pattern of the vortices, as well as the pressure distribution, exhibit excellent agreement with the featflow result.

\begin{figure}[!htbp]
	\centering
	\subfloat[Pressure Difference]{
	\includegraphics[width=0.5\textwidth]{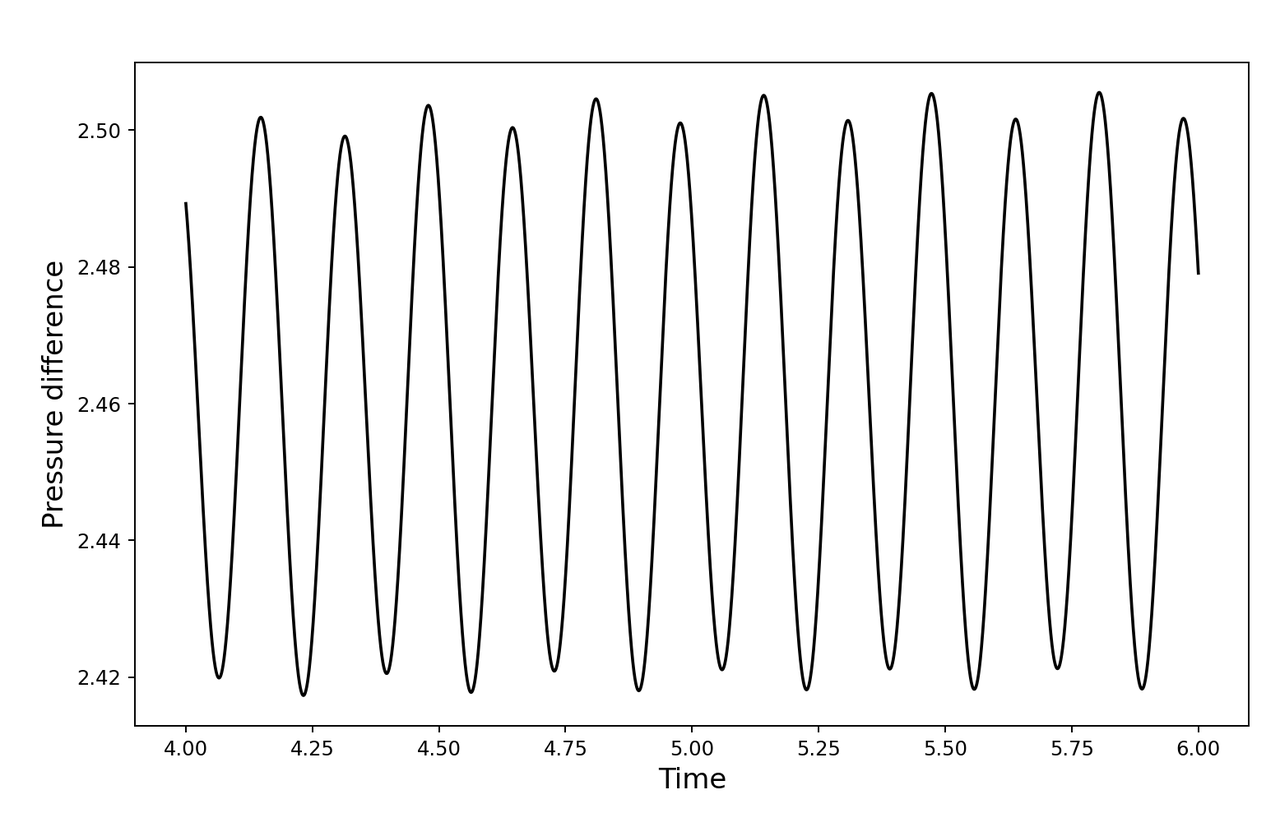}
	\label{}
	}
	\subfloat[Lift coefficient]{
		\includegraphics[width=0.5\textwidth]{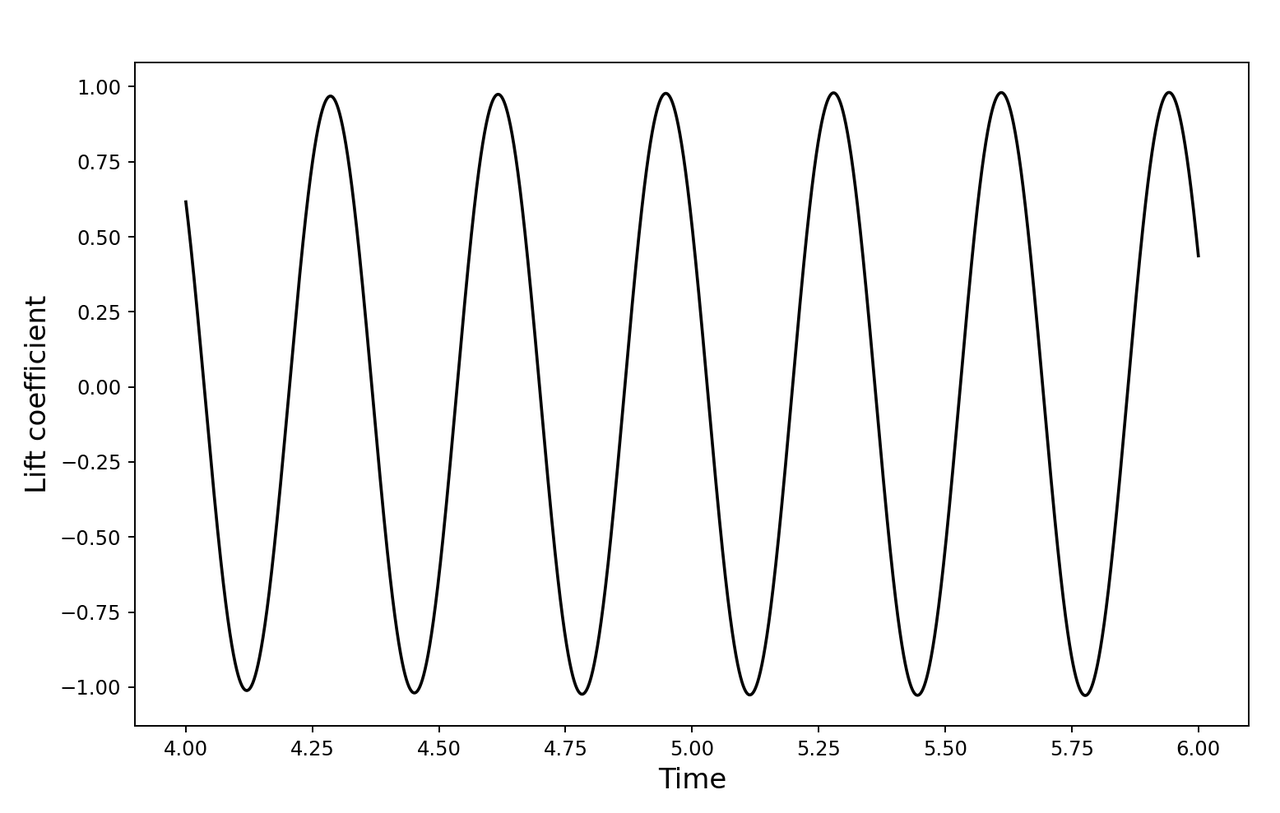}
		\label{}
	}
	\caption{Time histories of the pressure difference and  the lift coefficient for the flow past a cylinder benchmark}
	\label{fig:flow_past_cylinder_data}
\end{figure}

For a more rigorous quantitative validation, the time histories of the lift coefficient ($C_L$) on the cylinder and the pressure drop between the upstream surface  and the downstream surface  of the cylinder were monitored. Figure \ref{fig:flow_past_cylinder_data} presents these results after the flow has reached a periodic steady state. 

Our results are in excellent quantitative agreement with the reference data from the literature \cite{JOHN}.For the lift coefficient , simulation exhibits clear periodic oscillations with an observed period of approximately 0.3 seconds and a stable amplitude of approximately $\pm 1.0$. The pressure difference also demonstrates regular periodic fluctuations, with a period of approximately 0.15 seconds. The mean value (oscillation center) of the pressure difference is approximately 2.46, and the fluctuation range is precisely captured within the [2.42, 2.51] interval.

Having rigorously validated the solver's numerical accuracy and its capability to capture unsteady vortex shedding using the classic DFG 2D-2 cylinder benchmark, we further demonstrate the FEALPy.CFD framework's ability to handle complex and arbitrary geometries. To this end, we simulate the flow past a standard airfoil (NACA 0012).

\begin{figure}[!htbp]
	\centering
	\includegraphics[width=0.85\textwidth]{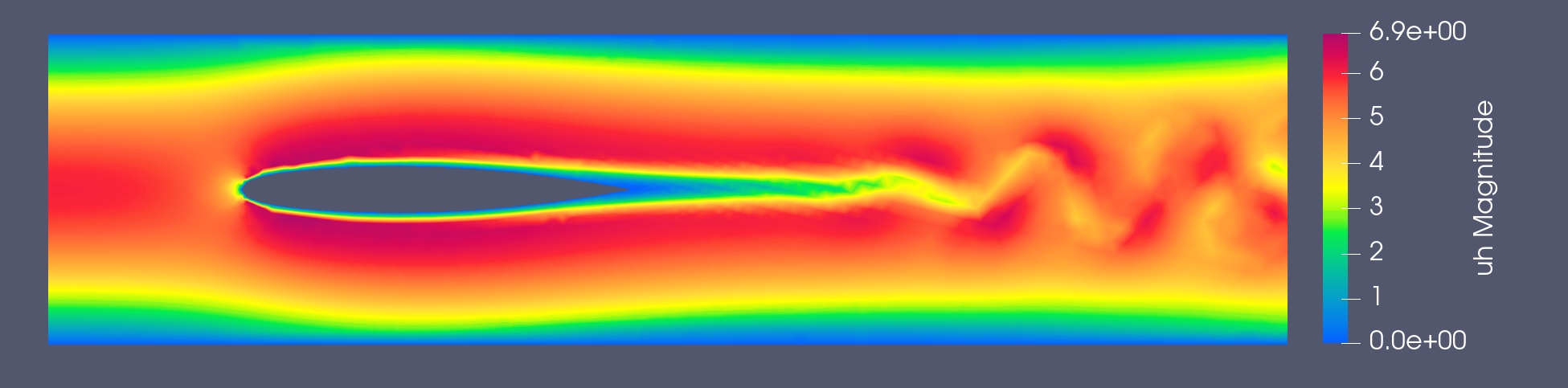}
	\caption{Flow around the airfoil}
	\label{airfoil}
\end{figure}

As shown in Figure \ref{airfoil}, FEALPy.CFD successfully captures the key physical phenomena of the flow around the airfoil, including the boundary layer development on the airfoil's surface, flow separation near the trailing edge, and the unsteady vortex shedding and evolution in the wake. This successful validation serves as a reliable foundation for its application in broader engineering contexts, such as fluid-structure interaction and aerodynamics.

\subsection{Taylor-Green Vortex}
To showcase the FEALPy.CFD framework's ability to integrate disparate numerical paradigms, this section employs the mesh-free SPH method to simulate the classic Taylor-Green Vortex (TGV) problem \cite{TAYLOR}. The selection of this benchmark serves a dual purpose: it not only validates the accuracy of the SPH solver in capturing transient vortex evolution but also highlights the framework's architectural flexibility as an open platform capable of supporting algorithms fundamentally different from traditional mesh-based methods.

The TGV problem is a canonical benchmark in fluid dynamics that describes the temporal evolution and subsequent viscous decay of a set of initially smooth, counter-rotating vortices within a periodic square domain. Its analytical solution is given by:
\begin{equation}
	\begin{aligned}
		u_1(x,y,t) & =-e^{-8\pi t}\cos(2\pi x)\sin(2\pi y) \\
		u_2(x,y,t) & =e^{-8\pi t}\sin(2\pi x)\cos(2\pi y)
	\end{aligned}
\end{equation}

\begin{figure}[!htbp]
	\centering
	\subfloat[t=0.08s]{
		\includegraphics[width=0.25\textwidth]{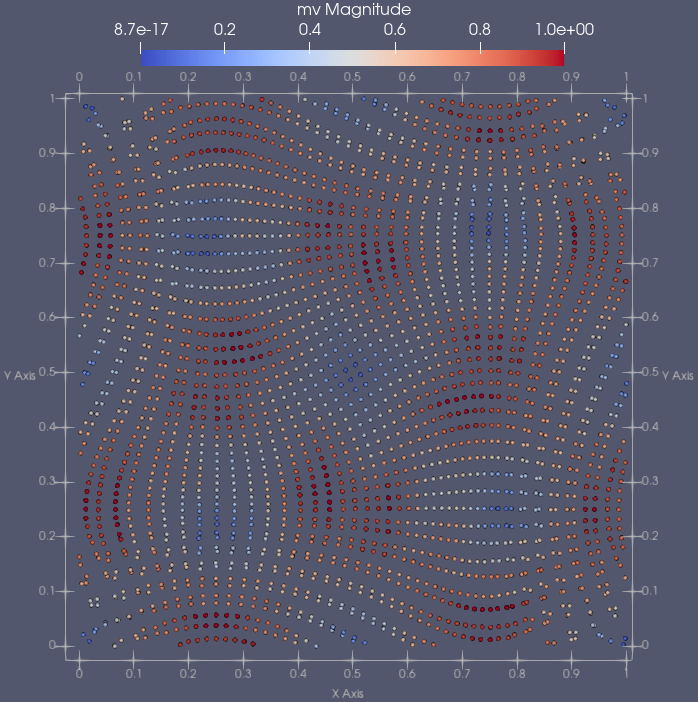}
		\label{}
	}
	\subfloat[t=0.28s]{
	\includegraphics[width=0.25\textwidth]{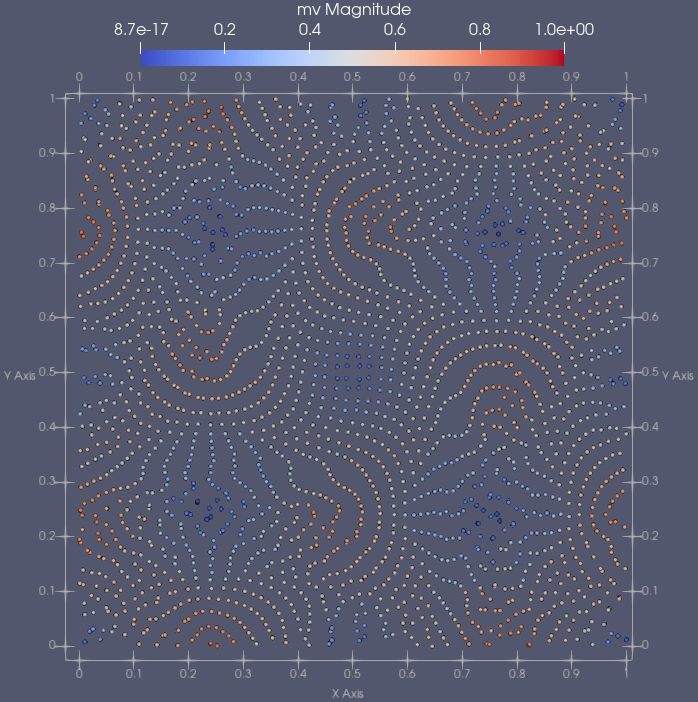}
	\label{}
	}
	\subfloat[t=0.48s]{
		\includegraphics[width=0.25\textwidth]{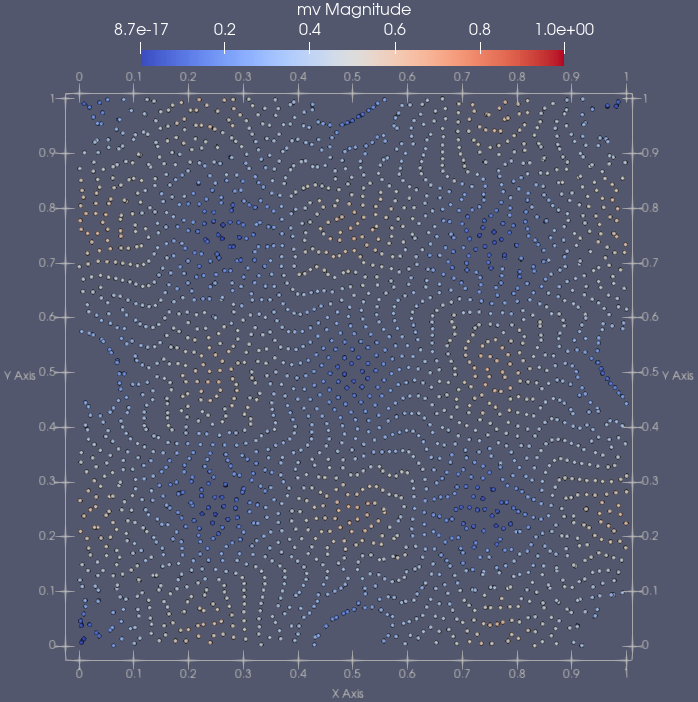}
		\label{}
	}
	\subfloat[t=0.68s]{
		\includegraphics[width=0.25\textwidth]{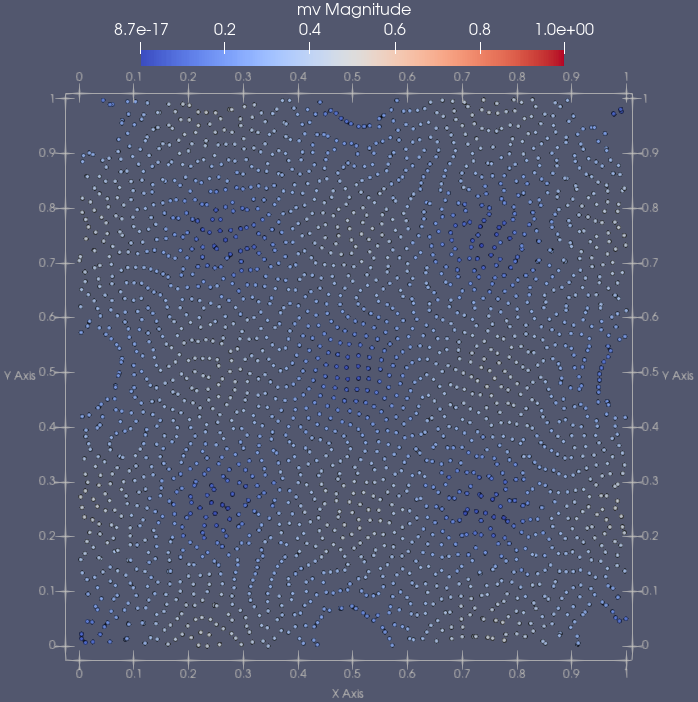}
		\label{}
	}
	\caption{Temporal evolution of the 2D Taylor-Green vortex at four different time instances.}
	\label{TGV}
\end{figure}

The Figure \ref{TGV} visually presents the evolution of the Taylor-Green vortex at different time instances, as simulated by the SPH method. It clearly illustrates the interaction of the initial four-vortex structure and its subsequent decay due to viscosity. As time progresses, the vortex intensity is observed to gradually diminish and the kinetic energy is dissipated, eventually leading the flow to a state of rest. This entire process is in excellent qualitative agreement with the theoretical behavior.

For a quantitative evaluation of the SPH solver's accuracy, a convergence study was performed by progressively refining the particle resolution. Table \ref{tab:particle_error} documents the errors at different particle counts, listing the Mean Square Error (MSE) for the individual velocity components, u and v, and the L2 norm of the error for the total velocity field.

\begin{table}[htbp]
	\centering
	\caption{}
	\label{tab:particle_error}
	
	\begin{tabular}{||c|c|c||}
		\hline
		Particles & MSE & $L_2$ error \\
		\hline
		2500   & 0.025358 & 0.096262 \\
		10000  & 0.003885 & 0.008149 \\
		40000  & 0.001313 & 0.004332 \\
		111000 & 0.000525 & 0.001502 \\
		\hline
	\end{tabular}
\end{table}
It is evident from the data in Table \ref{tab:particle_error} that all error metrics decrease systematically as the number of particles is increased (i.e., as the particle spacing is refined). This trend provides clear evidence that the implemented SPH algorithm exhibits the expected convergence behavior.

\subsection{Multi-backend Switching}
A core advantage of the FEALPy.CFD framework is its flexible multi-backend architecture, which allows users to seamlessly switch between computational backends like NumPy, PyTorch, and JAX according to the problem scale and available hardware. This section provides a quantitative performance evaluation of these backends. Using the Taylor-Green vortex and the flow past a cylinder as benchmark cases, we measure and compare their computational efficiency to demonstrate the significant performance gains enabled by this design.The system test environment equipped with an Intel® Core™ i7-8750H CPU, and an NVIDIA GeForce GTX 1050 GPU."

The backend can be switched by the user with a simple function call to set\_backend in the TensorBackendManager:
\begin{lstlisting}[language=python, label={code:SC}]
bm.set_backend('numpy')    # 切换到 NumPy 后端
bm.set_backend('pytorch')  # 切换到 PyTorch 后端
bm.set_backend('jax')      # 切换到 JAX 后端
\end{lstlisting}

Beyond the flexibility of multi-backend switching, FEALPy.CFD also leverages GPU acceleration to further boost computational efficiency, particularly when tackling large-scale problems. While computations are executed on the CPU by default (device='cpu'), users can easily migrate the task to a GPU (such as a CUDA-enabled device) as shown below:

\begin{lstlisting}[language=python,label={code:SC}]
bm.set_default_device('cuda')
\end{lstlisting}

This action ensures that all subsequently created tensors are allocated on the GPU by default. Consequently, all associated computations—such as matrix assembly and the solving of linear systems—are performed on the GPU.

We begin by employing the flow past a cylinder benchmark. The problem size, measured by the number of DoFs, is increased by progressively refining the mesh. We then measure and record the time for two key computational stages—matrix assembly and linear system solving—using the PyTorch backend on both CPU and GPU platforms.

\begin{table}[htbp]
	\centering
	\caption{Time (Assembly vs.Solve) for the flow past a cylinder benchmark using the fealpy.torch backend on CPU and GPU, for varying problem sizes}
	\label{tab:performance}
	\begin{tabular}{||c|c|c|c|c||}
		\hline
		\multirow{2}{*}{\textbf{DoFs}} & \multicolumn{2}{c|}{\textbf{fealpy.torch (GPU)}} & \multicolumn{2}{c||}{\textbf{fealpy.torch (CPU)}} \\
		\cline{2-5}
		& \textbf{Assembly} & \textbf{Solve} & \textbf{Assembly} & \textbf{Solve} \\
		\hline
		4770 & 557.976 [ms] & 0.605 [s] & 397.224 [ms] & 0.188 [s] \\
		\hline
		7584 & 590.333 [ms] & 0.639 [s] & 541.275 [ms] & 0.212 [s] \\
		\hline
		16204 & 589.038 [ms] & 0.641 [s] & 602.935 [ms] & 1.382 [s] \\
		\hline
		61651 & 666.251 [ms] & 1.096 [s] & 721.573 [ms] & 7.593[s] \\
		\hline
		95993 & 692.084 [ms] & 1.477 [s] & 699.163 [ms] & 15.181 [s] \\
		\hline
		170488 & 810.755 [ms] & 2.545 [s] & 812.155 [ms] & 30.848 [s] \\
		\hline
		245142 & 802.075 [ms] & 3.766 [s] & 1.373 [s] & 43.488 [s] \\
		\hline
		383034 & 831.198 [ms] & 6.393 [s] & 1.403 [s] & 83.328 [s] \\
		\hline
	\end{tabular}
\end{table}

The results presented in Table \ref{tab:performance} reveal distinct performance characteristics for each computational stage.

Matrix Assembly Stage: For smaller problem sizes, the CPU exhibits an advantage in assembly speed, likely due to the overhead of CPU-GPU data transfer.However, this trend reverses as the problem size grows. At the largest scale reported in Table \ref{tab:performance}, the massive parallel processing power of the GPU becomes dominant, with its assembly speed (831.198 ms) surpassing the CPU's (1.403 s) by a factor of approximately 1.7x.This performance gap widens significantly at extreme scales; a test case with 1.52 million DoFs was also evaluated , where the GPU assembly (1.187 s) was 4.2x faster than the CPU's (5.025 s).

Solve Stage: In this stage, which is typically the most computationally expensive, the GPU's performance advantage is substantial. The speedup effect becomes more pronounced as the problem size grows. For the largest case tested, the GPU is approximately 13 times faster than the CPU (6.393s vs. 83.328s). 

This result clearly demonstrates that the FEALPy.CFD's GPU backend can effectively accelerate the primary computational bottleneck in large-scale CFD simulations.

Subsequently, the transient Taylor-Green vortex problem is used to assess the overall performance of the various backends during long-time integration. Starting from identical initial conditions, we measure the total time required to execute simulations for 100, 1,000, and 10,000 time steps. This performance benchmark encompasses a comprehensive set of configurations: the NumPy backend on the CPU; the PyTorch backend on both CPU and GPU; and the JAX backend on the CPU, with its JIT compilation feature both enabled and disabled.

\begin{table}[htbp]
	\centering
	\caption{Overall performance comparison of the multi-backend architecture for the long-time integration of the Taylor-Green vortex problem.}
	\label{tab:backend_performance}
	
	\begin{tabular}{||c|c|c|c||}
		\hline
		Backend & 100 steps & 1000 steps & 10000 steps \\
		\hline
		JAX (CPU without JIT) & 44.76 s & 451.76 s & 5797.53 s \\
		JAX (CPU with JIT)    & 1.75 s  & 13.77 s  & 137.06 s  \\
		PyTorch (CPU)         & 16.73 s & 87.45 s  & 691.87 s  \\
		PyTorch (GPU)         & 12.78 s & 118.51 s & 1187.27 s \\
		NumPy                 & 22.10 s & 400.55 s & 5019.26 s \\
		\hline
	\end{tabular}
\end{table}

Analysis of Table \ref{tab:backend_performance} reveals two core characteristics of the FEALPy.CFD backend architecture. The most significant finding is the dramatic performance leap provided by JIT compilation: in the 10,000-step case, the JIT-enabled JAX backend was approximately 42 times faster than its non-JIT counterpart, establishing it as the clear performance leader across all configurations and proving  strong evidence that JIT (Just-In-Time) compilation is highly effective at mitigating the interpreter overhead inherent in Python, which is critical for loop-intensive scientific computing tasks. Secondly, the data reveals a subtle CPU/GPU performance trade-off: with the PyTorch backend, the GPU was faster for short runs (100 steps), but the CPU version became more efficient for longer simulations (1,000+ steps). This is likely because the computational workload per step was small, causing the cumulative overhead of CPU-GPU data transfers to eventually outweigh the GPU's parallel processing benefits in long-duration tasks.

In summary, the multi-backend architecture of FEALPy.CFD provides users with a powerful and flexible toolset for performance optimization. Whether leveraging JIT compilation to accelerate complex algorithms or harnessing GPU acceleration for large-scale problems, users can select the optimal computational strategy for their specific needs, all within a unified and consistent interface.

\subsection{Non-Newtonian Thermo-fluid Flow in a Planar Contraction}
To comprehensively demonstrate the integrated capabilities of FEALPy.CFD in handling complex industrial-type problems, this section simulates the classic problem of flow through a heated contracting channel. This benchmark is particularly challenging as it involves the strong coupling of fluid dynamics and heat transfer, while also incorporating a non-Newtonian power-law fluid model. The objective is to simultaneously validate the framework's accuracy and robustness on two distinct fronts: multiphysics coupling and support for complex constitutive models.

The numerical setup for this experiment references the 4:1 planar contraction flow benchmark from the literature\cite{Han}. The problem requires the coupled solution of the incompressible NS equations, the energy equation, and the power-law constitutive model. Key dimensionless parameters—such as the Reynolds number, Prandtl number, and the power-law index—are matched with the reference study to ensure the comparability of our results.

\begin{figure}[!htbp]
	\centering
	\subfloat[Coarse mesh]{
		\includegraphics[width=0.5\textwidth]{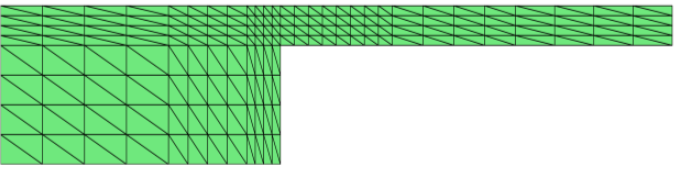}
		\label{}
	}
	\subfloat[Fine mesh]{
		\includegraphics[width=0.5\textwidth]{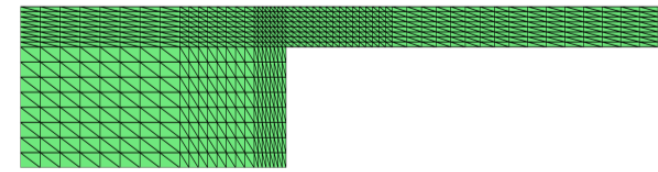}
		\label{}
	}
	\\
	\subfloat[Temperature field (coarse mesh)]{
		\includegraphics[width=0.5\textwidth]{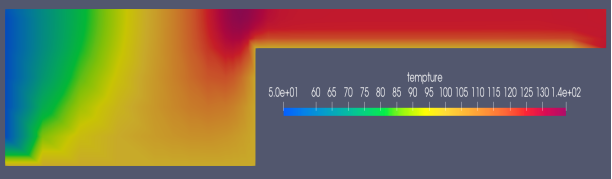}
		\label{}
	}
	\subfloat[Temperature field (fine mesh)]{
		\includegraphics[width=0.5\textwidth]{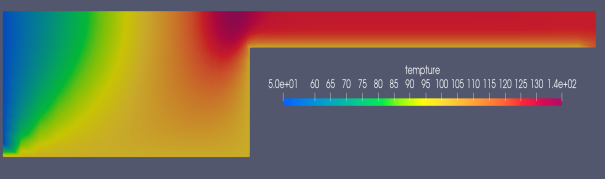}
		\label{}
	}
	\caption{Non-Newtonian flow in a heated 4:1 planar contraction: Mesh details and temperature comparison on coarse vs. fine grids}
	\label{Non-Newtonian flow}
\end{figure}

\begin{figure}[!htbp]
	\centering
	\includegraphics[width=0.9\textwidth]{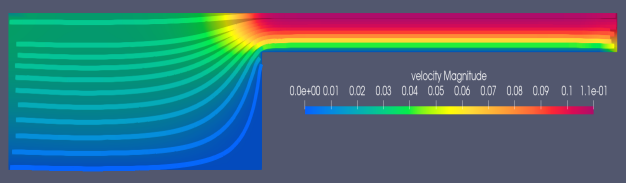}
	\caption{Streamline for the heated non-Newtonian contraction flow.}
	\label{streamlines non-Newtonian contraction flow}
\end{figure}

Figure \ref{Non-Newtonian flow} presents the results of the multiphysics coupling. It clearly depicts the temperature field distribution, illustrating the combined processes of heat convection and diffusion from the hot top wall into the fluid, particularly highlighting the formation of thermal boundary layers near the wall. To assess the reliability of the computation, we compared the results obtained on a coarse mesh  and a fine mesh . As observed, while finer details of the temperature field (particularly the gradients within the boundary layer) are more accurately resolved with mesh refinement, the overall macroscopic distribution remains consistent. This visually demonstrates the numerical convergence and robustness of FEALPy.CFD when handling strongly coupled fluid-thermal problems.

Figure \ref{streamlines non-Newtonian contraction flow} further provides a qualitative comparison between the computed streamlines and the reference results from the literature \cite{Han}. The overall distribution of velocity streamlines in the main flow region, including the fluid acceleration past the contraction, exhibits excellent consistency with the reference results. This validates the framework's capability to correctly describe the core behaviors of non-Newtonian hydrodynamics.

\subsection{The Rayleigh-Taylor Instability}
To demonstrate the FEALPy.CFD framework's capability for capturing complex interfacial dynamics, this part presents a simulation of the classic Rayleigh-Taylor (RT) instability, achieved by coupling the framework's NS solver with a phase-field model.

The RT instability occurs when a heavier fluid is situated above a lighter fluid within a gravitational field; under this condition, any small perturbation at the interface tends to grow over time, leading to the interpenetration of the two fluids.All parameters for this simulation reference the benchmark case presented by Lee \cite{LEE}.The density ratio of the heavy fluid to the light fluid is set at 3:1, corresponding to an Atwood number of $At=0.5$.

\begin{figure}[!htbp]
	\centering
	\includegraphics[width=0.9\textwidth]{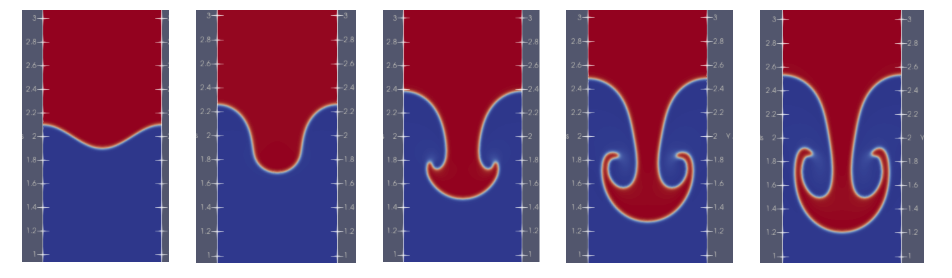}
	\caption{Simulation of the RT instability using the phase-field model coupled with the NS solver in FEALPy.CFD.}
	\label{RT}
\end{figure}
Figure \ref{RT} visually demonstrates the capability of the FEALPy.CFD framework, coupling its NS solver with a phase-field model, to successfully simulate RT instability. It is clearly observed that the results from FEALPy.CFD exhibit excellent agreement with the reference\cite{LEE} throughout the temporal evolution sequence, accurately capturing the key physical features characteristic of the RT instability: from the initial exponential growth of the small perturbation, to the distinct morphology of the descending "spikes" of heavy fluid and ascending "bubbles" of light fluid, evolving into the classic "mushroom-shaped" structures in later stages, and even capturing the secondary Kelvin-Helmholtz instabilities induced by high shear along the sides of the spikes, leading to characteristic vortical roll-ups. This high degree of agreement in interfacial morphology, evolution rate, and the capture of secondary instabilities strongly validates the ability of the FEALPy.CFD framework and its integrated phase-field model to accurately and robustly handle complex interfacial dynamics.

\subsection{Simulation of the "Tiger Stripe" Phenomenon}
Building upon the framework's previously validated capabilities in multiphysics (Section 5.5) and interface capturing (Section 5.6), we now extend its application to address a significant industrial challenge: the 'tiger stripe' defect commonly encountered in injection molding. 

This phenomenon, a critical surface quality issue often referred to as "flow marks," is characterized by alternating transverse bands perpendicular to the flow direction, typically arising from flow front instabilities occurring between dissimilar polymer melts. The precise physical mechanism is complex and still a subject of research, with leading hypotheses attributing the defect to a "slip-stick" phenomenon at the mold wall\cite{MATHIEU}. 

\begin{figure}[!htbp]
	\centering
	\subfloat[experimental phenomenon]{
		\includegraphics[width=0.35\textwidth]{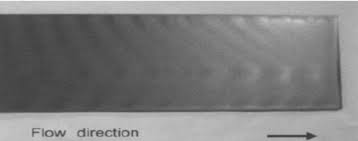}
		\label{tiger experimental}
	}
	\subfloat[FEALPy.CFD result]{
		\includegraphics[width=0.45\textwidth]{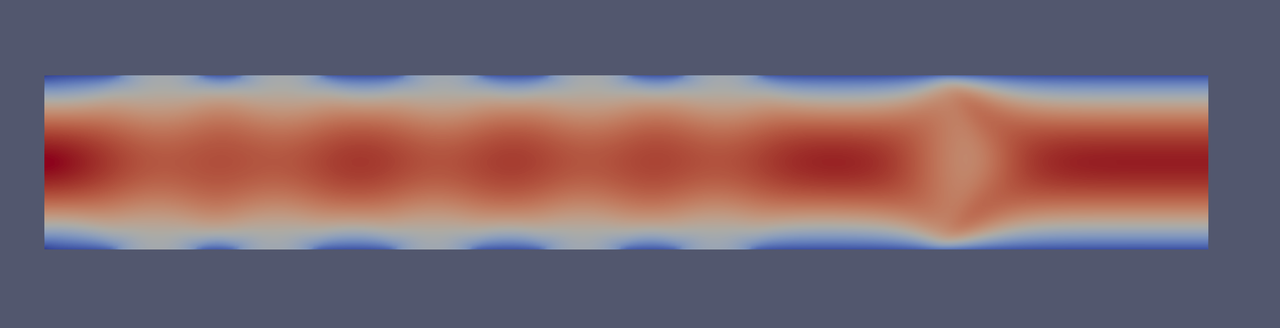}
		\label{tiger result}
	}
	\caption{Numerical simulation of the flow-front instability using FEALPy.CFD, capturing the characteristic alternating pattern consistent with the experimental phenomenon.}
	\label{tiger}
\end{figure}

This simulation couples the Cross-WLF constitutive model with a phase-field model to capture the unstable interfacial dynamics between two polymer melts during a rectangular mold filling process. Figure \ref{tiger} offers a direct qualitative comparison between experimental observations Figure \ref{tiger experimental} and our numerical simulation results Figure \ref{tiger result}. The simulation qualitatively reproduces the characteristic periodic transverse ripple pattern observed experimentally.

This successful reproduction of a complex non-Newtonian multiphase flow instability validates the potential of FEALPy.CFD as a robust tool for diagnosing and optimizing industrial manufacturing processes.

\section{Conclusion}
This paper introduces FEALPy.CFD, a high-performance, open-source framework designed to address critical challenges in the CFD domain regarding usability, reliance on single discretization methods, and hardware performance portability. Core innovations—validated through a series of tests—include a unified architecture integrating mesh-based (FEM, FVM) and mesh-free (SPH) methods, a usability-focused dual-layer API catering to both users and researchers, and a backend-agnostic design (NumPy, PyTorch, JAX). Performance benchmarks confirm the architecture's power: JAX's JIT compilation delivers over 40x speedup, while GPU acceleration provides an order-of-magnitude performance boost for large-scale solves. Future work will concentrate on introducing turbulence models (e.g., LES/RANS) and integrating support for more emerging computational backends (e.g., TensorFlow). Consequently, FEALPy.CFD offers a modern, flexible alternative that combines algorithmic comprehensiveness with hardware-agnostic high performance, lowering the barrier to complex simulations.
%%%% Acknowledgments %%%%%%%%
\section*{Acknowledgments}
The first and third authors were supported by the National Natural Science Foundation of China (NSFC) (Grant Nos. 12371410, 12261131501) and the National Key R\&D Program of China (Grant No. 2024YFA1012600). All authors were supported by the Kingfa Science and Technology Company.

\end{document}